%% file: cdm-buildup-imf.tex
\documentclass[iop]{emulateapj}

\usepackage[T1]{fontenc}
\usepackage{ae,aecompl}

\usepackage{graphicx}
\usepackage{amsmath}
\usepackage{amssymb}
\usepackage{enumitem}

\providecommand{\e}[1]{\ensuremath{\times 10^{#1}}}
\newcommand{\RN}[1]{%
  \textup{\uppercase\expandafter{\romannumeral#1}}%
}

\usepackage{hyperref}

\hypersetup{
    colorlinks,
    citecolor=cyan,
    filecolor=cyan,
    linkcolor=cyan,
    urlcolor=cyan
}

\hypersetup{urlcolor=cyan}

\usepackage{ulem}

\slugcomment{Draft version June 1, 2017}

\shorttitle{$\Lambda$CDM buildup and IMF variations}

\shortauthors{Blancato, Genel, \& Bryan}

\begin{document}

\title{Implications of galaxy buildup for putative IMF variations in massive galaxies}

\author{Kirsten Blancato\altaffilmark{1}\dag, Shy Genel\altaffilmark{2,1}, Greg Bryan\altaffilmark{1,2} }

\altaffiltext{1}{Department of Astronomy, Columbia University, 550 West 120th Street, New York, NY 10027}
\altaffiltext{2}{Center for Computational Astrophysics, Flatiron Institute, 162 Fifth Avenue, New York, NY 10010}
\altaffiltext{\dag}{Electronic address: knb2128@columbia.edu}

%%%%%%%%%%%%%%%%%%% ABSTRACT %%%%%%%%%%%%%%%%%%%
\begin{abstract}

\noindent Recent observational evidence for initial mass function (IMF) variations in massive quiescent galaxies at $z = 0$ challenges the long-established paradigm of a universal IMF. While a few theoretical models relate the IMF to birth cloud conditions, the physical driver underlying these putative IMF variations is still largely unclear. Here we use post-processing analysis of the Illustris cosmological hydrodynamical simulation to investigate possible physical origins of IMF variability with galactic properties. We do so by tagging stellar particles in the simulation (each representing a stellar population of $\approx10^{6}~\mathrm{M}_{\odot}$) with individual IMFs that depend on various physical conditions, such as velocity dispersion, metallicity, or SFR, at the time and place the stars are formed. We then follow the assembly of these populations throughout cosmic time, and reconstruct the overall IMF of each $z=0$ galaxy from the many distinct IMFs it is comprised of. Our main result is that applying the observed relations between IMF and galactic properties to the conditions at the star-formation sites does not result in strong enough IMF variations between $z = 0$ galaxies. Steeper physical IMF relations are required for reproducing the observed IMF trends, and some stellar populations must form with more extreme IMFs than those observed. The origin of this result is the hierarchical nature of massive galaxy assembly, and it has implications for the reliability of the strong observed trends, for the ability of cosmological simulations to capture certain physical conditions in galaxies, and for theories of star-formation aiming to explain the physical origin of a variable IMF.
\end{abstract}


%%%%%%%%%%%%%%%%% INTRODUCTION %%%%%%%%%%%%%%%%%%

\section{Introduction}
\label{sec:introduction}

The stellar initial mass function (IMF) has long been thought a universal feature of star formation. Although observational support for a universal IMF has been mounting for more than half a century, the physical theory behind the IMF still remains an unsolved problem in star formation physics \citep[e.g.][]{imf-bastian}. The elusive origin of the IMF stems from deficient knowledge of the physics governing how molecular clouds collapse and fragment to form protostars and the subsequent accretion of gas onto these protostars \citep[e.g.][]{imf-offner}. Observational measurements of the IMF have also proved to be particularly challenging, both for resolved stellar populations where individual stars can be counted, and for distant \citep[e.g.][]{imf-chabrier-2}, unresolved stellar populations where the IMF must be inferred more indirectly \citep[e.g.][]{imf-tang}. Recent observations suggesting a variable IMF, as measured within massive elliptical galaxies, serve to further compound the elusive origin of the IMF \citep[e.g.][]{imf-cappellari-1,imf-conroy-1,imf-spiniello,imf-ferreras-2}. While understanding the shape of the IMF would certainly provide insight into the physical processes that control star formation, the IMF also has vast implications for the study of galaxy populations as the IMF influences nearly all observable galaxy properties including stellar mass, luminosity, metal content, and star formation history.

Observations of stellar populations within and near the Milky Way gave rise to the conception of a universal IMF. The first determination of the IMF was by \cite{imf-salpeter}
who, using field stars in the solar neighborhood, found the IMF to follow a unimodal power law with a slope of $x = 2.35$. Since Salpeter's initial measurement the IMF has been extensively measured in nearby, resolved stellar populations. These measurements have revealed the IMF to turn over at lower masses, following a \cite{imf-chabrier} log-normal or a \cite{imf-kroupa} segmented power law IMF below $\sim$1 M$_{\odot}$ and the original Salpeter IMF at stellar masses greater than $\sim$1 M$_{\odot}$. The shape of the IMF has been measured to be largely consistent across a variety of stellar populations within and near the Milky Way, including young clusters \citep[e.g.][]{imf-hillenbrand,imf-weights}, open clusters \citep[e.g.][]{imf-massey,imf-carraro}, globular clusters \citep[e.g.][]{imf-mclaughlin,imf-kruijssen}, the Large Magellanic Cloud (LMC) \citep[e.g.][]{imf-kerber,imf-dario} and Small Magellanic Cloud (SMC) \citep[e.g.][]{imf-sirianni,imf-sabbi}, and M31 and M32 \citep[e.g.][]{imf-zieleniewski}. For a detailed discussion of the nuances of local IMF observations we direct the reader to \cite{imf-chabrier-2} and \cite{imf-bastian}.

Surprisingly, observations over the past few years of more distant stellar populations suggest deviations from the universal IMF inferred in the Local Group. Several independent methods have been implemented to study the IMF in unresolved stellar populations, including: (1) dynamical studies where stellar population synthesis (SPS) mass-to-light ratios are compared to dynamically derived mass-to-light ratios \citep[e.g.][]{imf-dutton,imf-cappellari-1,imf-cappellari-2,imf-conroy-1,imf-conroy-2,imf-martin-1}, (2) absorption line studies where spectral features either sensitive or anti-sensitive to dwarf stars provide the constraints \citep[e.g.][]{imf-cenarro,imf-vandokkum-1,imf-ferreras-3,imf-labarbera-1,imf-spiniello,imf-mcconnell}, and (3) lensing studies where SPS masses are compared to masses derived from gravitational lensing \citep[e.g.][]{imf-ferreras-1,imf-ferreras-2,imf-auger,imf-treu,imf-thomas,imf-posacki,imf-leier}. 

These measurements predominately infer that the IMF of nearby early-type galaxies (ETGs) becomes more bottom-heavy, i.e.~having increasingly numerous low-mass stars with respect to high-mass stars, at higher values of galaxy properties such as velocity dispersion ($\sigma$), metallicity ([M/H]), and metal abundance ratio ([$\alpha$/Fe]). For example, \cite{imf-conroy-2} find IMF variations with velocity dispersion for a sample of compact ETGs by using mass-to-light ratio as a proxy for the fraction of low-mass stars, as these systems are believed to be stellar dominated at their centers. They find low velocity dispersion galaxies ($\sigma$ $\sim$ 100 km s$^{-1}$) to be best described by a Milky Way-like IMF, galaxies with intermediate velocity dispersions ($\sigma$ $\sim$ 160 km s$^{-1}$) best fit by a Salpeter IMF, and galaxies with $\sigma$ $\sim$ 250 - 300 km s$^{-1}$ best described by an IMF even more bottom-heavy than the Salpeter IMF. Similar IMF trends have also been observed to scale with metallicity. For example, using a sample of ETGs from the CALIFA survey, \cite{imf-martin-1} find the most metal-poor ETGs in their sample ([M/H] $\sim$ -0.2) to be best described by an IMF slope of $x$ $\sim$ 2 and the most metal-rich ETGs ([M/H] $\sim$ 0.2) best described by an IMF slope of $x$ $\sim$ 2.9. 

IMF variations have also been observed \textit{within} galaxies, highlighting the complications of systematics like aperture radius when comparing IMF measurements across studies \cite[e.g.][]{imf-martin-2, imf-lb, imf-vd} (however, see \cite{imf-vaughan} who find a constant IMF at all radii for two ETGs with $\sigma$ = 410 and 260 km s$^{-1}$). In particular, using deep spectroscopic data taken at various fractions of the effective radius ($R_{e}$), \cite{imf-martin-2} find significant radial IMF trends for the highest velocity dispersion galaxies in their sample ($\sigma$ $\sim$ 300 km s$^{-1}$), starting with an IMF slope of $x$ $\sim$ 3 at galaxy centers, down to an IMF slope of $x$ $\sim$ 1.9 at $r = 0.7 ~\mathrm{R}_{e}$.  However, for the lower velocity dispersion ETG in their sample ($\sigma$ $\sim$ 100 km s$^{-1}$) the IMF is found to be constant with galactocentric distance. These trends may be understood in the context of the `minor mergers' scenario \citep{imf-naab,imf-oser,r-gomez}, according to which the most massive ETGs accrete a large number of small galaxies in particular in their outer regions.

Similar IMF variations have also been recently reported for high redshift ETGs. Comparing dynamical to SSP masses, \cite{imf-shetty-1, imf-shetty-2} find a Salpeter IMF, rather than the `universal' Chabrier, for their sample of massive (\textgreater ~10$^{11}$ M$_{\odot}$) ETGs at $z \sim 0.8$. For a sample of ETGs at $z \sim 1.4$ with both dynamical and photometric mass estimates, \cite{imf-gargiulo-2} report an IMF-$\sigma$ relation consistent with the trends observed at $z \sim 0$. They further posit that the IMF of dense (\textgreater ~2500 M$_{\odot}$ pc$^{-2}$)  ETGs is independent of redshift over the past $\sim$9 Gyr. Additionally, based on the dwarf-sensitive TiO$_{2}$ feature, \cite{imf-martin-4} find similar IMF variations for massive ETGs from 0.9 \textless ~$z$ \textless ~1.5 that are consistent with a constant IMF over the past $\sim$8 Gyr.

Though the evidence for IMF variations is mounting, a consensus has yet to be reached, as several studies report discrepant results. For example, \cite{imf-conroy-1} find for their sample of ETGs the strongest IMF correlation to be with [Mg/Fe], with weaker IMF-$\sigma$ and IMF-[M/H] correlations. \cite{imf-labarbera-2}, on the other hand, report negligible IMF correlation with [Mg/Fe]. Additionally, based on studies of the low-mass X-ray binaries (LMXB) of ETGs, whose number is expected to scale with the IMF, \cite{imf-peacock} argue for an invariant IMF reporting a LMXB population per mass that is constant across a range of galaxy velocity dispersions. Studies based on lensed galaxies have also yielded discrepant results. \cite{imf-smith-1} found that for two strong lens ETGs in the SINFONI Nearby Elliptical Lens Locator Survey (SNELLS) with $\sigma$ $\sim$ 300 km s$^{-1}$ a bottom-heavy IMF is ruled out in favor of a Kroupa IMF, but find their 1.14 Na \RN{1} $\mu$m index strengths to suggest they have bottom-heavy IMFs \citep{imf-smith-2}. Recently \cite{imf-newman} compared lensing, dynamical, and SPS techniques for inferring the IMF of these SNELLS galaxies, finding that the SPS stellar mass-to-light ratios exceed the total lens mass-to-light ratio, and that there is even a significant discrepancy between the lensing and dynamical masses. \cite{imf-newman} discusses several possibilities for the origin of these tensions, but this study suggests there could be systematic errors in at least one of the techniques used to probe the IMF of ETGs.

These discrepancies highlight the importance of understanding the uncertainties in inferring IMF variations. Emphasizing the difficulties of inferring the IMF from integrated light, \cite{imf-tang} find that the degeneracy between a bottom-heavy IMF and decreasing AGB (asymptotic giant branch) strength is only confidently broken for old, metal-rich galaxies with a combination of accurate spectra and photometric observations at the .02 mag level. \cite{imf-clauwens-2} examines the influence of measurement error and selection bias on IMF variations using a sample of galaxies from the ATLAS-3D project. They find that $\sim$30\% gaussian errors on kinematic measurements of mass-to-light ratios lead to similar IMF variations as reported in \cite{imf-cappellari-1}, emphasizing the importance of correctly modeling measurement errors. \cite{imf-clauwens-2} also find that galaxy selection can significantly influence the inferred IMF trend. Placing a cut on star-formation (as most studies reporting IMF variations do) removes low velocity dispersion galaxies with IMFs comparable to quiescent, high velocity dispersion galaxies. Additionally, excluding galaxies with kinematic masses below the ATLAS-3D mass completeness threshold (2\e{10} M$_{\odot}$) removes the IMF trend with velocity dispersion.

If however proven robust, a variable IMF will have significant implications for our current understanding of galaxy formation and evolution: an understanding that has largely been developed under the assumption of a universal IMF. In particular, a variable IMF will affect the derived properties of galaxy populations. For example, \cite{imf-clauwens} quantifies the effect of the metallicity dependent IMF relation in \cite{imf-martin-1} on the derived quantities of nearly $2\times10^{5}$ SDSS galaxies. Inferred star formation rates increase by up to two orders of magnitude and stellar mass densities increase by a factor of 2.3 compared to what is inferred with the Chabrier IMF. Adopting an IMF relation dependent on velocity dispersion, \cite{imf-mcgee} find similar effects on the derived properties of SDSS galaxies, with the shape of the high-mass end of the galaxy stellar mass function shifting from the familiar exponential (using a Chabrier IMF) to a power-law. From a more theoretical standpoint, \cite{imf-vincenzo} use a numerical chemical evolution code to quantify the effect of the IMF shape on metal yields of galaxies. Exploring both the upper mass cut off and the slope of the IMF, they find that the metal yield can vary by up to a full order of magnitude.

Against this observational background, several star-formation models have been recently developed that predict IMF variations. Some models find that the larger density fluctuations associated with higher Mach numbers in star-forming disks cause the low-mass turnover of the pre-stellar core mass function (CMF) to shift to lower masses, implying a more bottom-heavy IMF (e.g.~\citealp{r-hopkins, imf-chabrier-3, imf-gus}; but see \citealp{imf-bertelli}). These models are able to reconcile the universal IMF found across a range of Milky Way stellar populations with a bottom-heavy IMF in more extreme star-formation environments such as starbursts. Also potentially able to account for differences between such environments is the \cite{imf-krumholz} derivation of the low-mass turnover as a function of fundamental constants and a weak dependence on interstellar pressure and metallicity. Another theory is IGIMF (Integrated Galaxy-wide stellar Initial Mass Function), which formulates the shape of the overall IMF of a galaxy based on the properties of its individual molecular clouds, which are in turn controlled by the total galaxy SFR \citep{imf-weidner-2}. In this theory, high SFR environments, such as starbursts, are predicted to undergo star-formation which follows a top-heavy IMF, so the excess mass inferred for massive elliptical galaxies is predicted to be in part due to stellar remnants.

\input{illustris-table}

Semi-analytical and numerical models have also recently been employed to study IMF variations. The impact of IMF variations on the chemical abundances of galaxies has been quantified using both \texttt{SAG} \citep{imf-gargiulo} and \texttt{GAEA} \citep{imf-fontanot}. \cite{imf-fontanot-2} uses \texttt{MORGANA} to investigate the effect of IMF variations on both stellar mass and star-formation rate, reporting that top-heavy IMF in high star-formation rate environments produces the largest deviations from properties derived under a standard IMF. A variable IMF has also been explored in the context of simulations of individual galaxies, finding that chemical evolution is dependent on the assumed IMF \citep{imf-bekki, imf-few}. Most recently, \cite{imf-sonnenfeld} developed a simple model of merger-driven galaxy evolution to predict the evolution of IMF trends with mass and velocity dispersion from $z=2$ to $z=0$, predicting that the IMF slope of a galaxy is steeper at earlier times.

In this study, we use the Illustris cosmological hydrodynamical simulation to connect the physical conditions in which stars form to the global properties of galaxies at $z = 0$. We construct the IMF for a sample of $z=0$ Illustris galaxies by using prescribed IMF relations applied to the birth properties of the individual stellar populations that comprise each galaxy. By attempting to reproduce observed relations between the overall IMF of a galaxy (or of its central parts) and its $z = 0$ properties, we are able to provide constrains on relations between IMF and physical conditions at the time of stellar birth.

This paper is organized as follows. In Section \ref{sec:methods} we describe the Illustris simulation, our galaxy selection, and the method for constructing the IMF. In Section \ref{sec:results1} we present the primary results: global IMF trends with galactic properties at $z=0$. In Section \ref{sec:results2} we explore additional constrains: radial trends, scatter, and redshift evolution. In Section \ref{sec:res_var}, we test the robustness of our results to both resolution and variations in the simulation. Finally, in Section \ref{sec:discussion} we discuss our findings, and summarize in Section \ref{sec:conclusion}.

%%%%%%%%%%%%%%%%% ILLUSTRIS %%%%%%%%%%%%%%%%%%
  
\section{Methods}
\label{sec:methods}

\subsection{Simulation suite}
\label{sec:illustris}

To investigate possible physical origins of the observed IMF variations we primarily use cosmological simulations from the Illustris Project \citep{ib-vogelsberger-a,ib-vogelsberger-b,ib-genel}, and in particular the highest resolution hydrodynamical simulation in the suite, the Illustris simulation. These simulations evolve down to $z=0$ a volume large enough to contain statistically significant galaxy populations, and incorporate crucial physics resulting in many realistic galaxy properties. Illustris has been used to study a diverse range of topics in galaxy evolution including, in particular, topics directly relevant to this work, such as the formation of massive, compact ETGs \citep{ib-wellons-1,ib-wellons-2} and the stellar mass assembly of galaxies \citep{r-gomez}.

The Illustris simulation treats hydrodynamical calculations using the moving-mesh code \texttt{AREPO} \citep{ib-springel}, which has proven advantages over both adaptive mesh refinement (AMR) and smoothed particle hydrodynamics (SPH) techniques \citep{ib-sijacki,ib-kere,ib-vogelsberger-2}.  Gravitational forces are computed using a Tree-PM technique \citep{ib-xu} that calculates short-range forces using the tree algorithm and long-range forces using the particle mesh (PM) method. A $\Lambda$CDM cosmology with $\Omega_{m} = 0.2726$, $\Omega_{\Lambda} = 0.7274$, $\Omega_{b} = 0.0456$, and $h = 0.704$ from WMAP9 \citep{ib-hinshaw} is adopted for all simulations used in this study. The galaxy formation physics implemented in Illustris includes radiative cooling, star formation and evolution, including chemical enrichment, black hole seeding and accretion, stellar feedback in the forms of ISM pressure and galactic winds, as well as AGN feedback. Since our study focuses on the stellar populations in Illustris, below we provide a short description of the stellar formation and evolution model.  For an in depth discussion of all of the physical models included in Illustris the reader is referred to \cite{ib-vogelsberger-3}.

Stellar particles form according to the Kennicutt-Schmidt relation \citep{ib-kennicutt} from dense ISM gas with a time scale of 2.2 Gyr at the density threshold of $n\approx0.13$ cm$^{-3}$. This gas is pressurized following an effective equation of state for a two-phase medium \citep{ib-springel-3}. Each stellar particle represents a simple stellar population (SSP) consisting of stars formed at the same time with the same metallicity. Stars in each stellar particle SSP return mass and metals to surrounding gas cells following their expected lifetimes with post-main sequence evolution occurring instantaneously, where low mass stars return mass through AGB winds and more massive stars return most of their mass to the ISM via supernovae. Hence, the mass loss and metal production of each stellar particle as a function of its age are calculated using tables in accordance with the particle's initial mass, metallicity, and assumed IMF.

The SSP of each stellar particle in Illustris is assumed to have a stellar mass distribution described by a Chabrier IMF. The IMF affects the mass and metal return via the evolution of high mass stars, as well as the energy available for galactic wind feedback, which however has a pre-factor that is a tunable parameter of the model. In Section \ref{sec:res_var} we explore simulations not included in the Illustris suite that are evolved with different IMFs, such as the Salpeter IMF as well as a variable IMF. In addition, we study simulations that adopt different degrees of feedback.
 
In post-processing, structure is identified in each snapshot first using the \texttt{FoF} (friends-of-friends) algorithm \citep{ib-davis} and then an updated version of the \texttt{SUBFIND} algorithm \citep{ib-springel-2,ib-dolag}. The \texttt{FoF} algorithm identifies dark matter halos using a linking length of one-fifth the mean separation between dark matter particles, with the baryonic particles (gas, stars, and black holes) assigned to the FoF group of their closest dark matter particle if it is close enough by the same separation criterion. The \texttt{SUBFIND} algorithm identifies gravitationally-bound substructure (subhalos) within each parent \texttt{FoF} group. The dark matter and baryonic components of each subhalo constitute what we refer to as a galaxy. 

The publicly released suite of hydrodynamical Illustris simulations\footnote{http://www.illustris-project.org/data/ \citep{ib-nelson}} includes three runs of the same volume at increasing resolution levels: Illustris-3,-2, and -1. Illustris-1 includes 1820$^{3}$ dark matter particles with masses $m_{\rm DM}$ = 6.26\e{6} M$_{\odot}$ and $\sim$1820$^{3}$ baryonic resolution elements with an average mass of $\overline{m_{b}}$ = 1.26\e{6} M$_{\odot}$, evolved within a (106.5 Mpc)$^{3}$ cube. More details about Illustris-1 and the other simulations used in this study are given in Table \ref{tab:illustris}.

\subsection{Galaxy selection}
\label{sec:selection}

In selecting galaxies in Illustris, we aim to mimic the typical properties of the ETGs examined in the observational IMF studies. We therefore first select galaxies with stellar masses greater than 10$^{10}$ M$_{\odot}$ and specific star formation rates \textless ~10$^{-11}$ yr$^{-1}$. The total stellar mass of each galaxy is calculated as the sum of the stellar particles assigned to it by \texttt{SUBFIND} and the specific star formation rate is calculated as the sum of the instantaneous SFRs of its gas cells, divided by the stellar mass. Both the stellar mass and sSFR are calculated within two times the stellar half-mass radius of each galaxy. For Illustris-1 at $z = 0$ these selection criteria result in a sample of 1160 galaxies. The mean stellar mass of this sample is $M_{*}$ = 10$^{10.88}$ M$_{\odot}$ and the mean specific star formation rate is sSFR = 2.45\e{-12} yr$^{-1}$. 

Additionally, as done in \cite{imf-spiniello}, we limit our sample to galaxies with stellar velocity dispersions greater than 150 km s$^{-1}$.The stellar velocity dispersion, $\sigma_{*}$, for each galaxy is calculated as the one-dimensional, $r$-band luminosity-weighted velocity dispersion of the stellar particles falling within one-half the projected\footnote{The projected half-mass radius $R^{p}_{1/2}$ is calculated as the radius containing half the total stellar mass of the galaxy, including stellar particles that fall within this projected radius as viewed from the $z$-direction.} stellar half-mass radius (0.5$R^{p}_{1/2}$). The velocity dispersion criterion reduces our sample from 1160 to 371 galaxies, with $\overline{M_{*}}$ = 10$^{11.4}$ M$_{\odot}$, $\overline{\mathrm{sSFR}}$ = 1.22\e{-12} yr$^{-1}$, and $\overline{\sigma_{*}}$ = 198 km s$^{-1}$.

\subsection{IMF construction}
\label{sec:imf}

Since the IMF is set at the time of stellar birth, we probe the birth conditions of the stellar particles belonging to our selection of $z = 0$ galaxies. To do this we trace each of the 196335880 stellar particles belonging to the 371 $z = 0$ selected galaxies back to the snapshot in which it first appears and compute several physical quantities (discussed in Section \ref{sec:results1}) that represent its birth conditions.

Following \cite{imf-spiniello,imf-treu} we assign an IMF mismatch parameter, $\alpha_{\rm IMF}$, to each stellar particle:
\begin{equation} \label{eq:alpha}
\alpha_{\rm IMF} = \frac{(M_{*}/L)}{(M_{*}/L)_{\rm Salp}}, 
\end{equation}

\medskip

\noindent where $(M_{*}/L)_{\rm Salp}$ is the mass-to-light ratio expected assuming a Salpeter IMF and $(M_{*}/L)$ is the actual mass-to-light ratio assumed for the particle. For stellar populations with an IMF more `bottom-heavy' compared to the Salpeter IMF, $\alpha_{\rm IMF}$ \textgreater ~1, while stellar populations `bottom-light' compared to the Salpeter IMF have $\alpha_{\rm IMF}$ \textless ~1. A Chabrier IMF is described by $\alpha_{\rm IMF}$ = 0.6.

To construct the overall $\alpha_{\rm IMF}$ of each galaxy described in Section \ref{sec:selection} we mass-weight $\alpha_{\rm IMF}$$^{-1}$ using the birth mass of all (or the innermost subset of) the stellar particles comprising a galaxy. This is equivalent to summing up the light, $L/L_{x=2.35}$, assigned to the stellar particles belonging to each galaxy.

Beginning our exploration, we are inspired by the observations when assigning an $\alpha_{\rm IMF}$ to the stellar particles. For example, we use the $\alpha_{\rm IMF}$-$\sigma_{*}$ relation presented in \cite{imf-spiniello} as an input relation applied to the local velocity dispersions of the stellar particles. In general though, we have the freedom to construct input $\alpha_{\rm IMF}$ relations that scale with various physical quantities at the star-formation sites. In Section \ref{sec:results1} we present the Illustris log($\alpha_{\rm IMF}$)-$\sigma_{*}$ relations constructed using five different physical quantities associated with the birth conditions of each stellar particle: global stellar velocity dispersion ($\sigma_{*}$), local dark matter velocity dispersion ($\sigma_{\mathrm{birth}}$), local metallicity ([M/H]), global star-forming gas velocity dispersion ($\sigma_{\mathrm{gas}}$), and global star-formation rate (SFR).

%%%%%%%%%%%%%%%%%%% RESULTS %%%%%%%%%%%%%%%%%%

\section{Investigations of IMF physical drivers}
\label{sec:results1}

\subsection{Global velocity dispersion}
\label{sec:globalstar}

We begin our investigation by first constructing the overall log($\alpha_{\rm IMF}$) of each galaxy in the Illustris sample based on the global stellar velocity dispersion, $\sigma_{*}$. We trace each stellar particle belonging to a $z=0$ galaxy back to the progenitor galaxy it was formed in, and compute that galaxy's stellar velocity dispersion in exactly the same way $\sigma_{*}$ was computed for the $z=0$ galaxy sample. 

With the global $\sigma_{*}$ associated with each star particle, we construct the overall IMF mismatch parameter (log($\alpha_{\rm IMF}$)) according to the prescription outlined in Section \ref{sec:imf}. As a starting point, we apply the observed relation presented in \cite{imf-spiniello},

\begin{equation} \label{eq:spiniello}
\mathrm{log}(\alpha_{\rm IMF}) = (1.05\pm.2) \mathrm{log}(\sigma_{*}) - (2.5\pm.4),
\end{equation}

\smallskip

\noindent which is derived over a range of SDSS ETGs ($z \le 0.05$) with velocity dispersions between $\sigma_{*}$ = 150 km s$^{-1}$ and $\sigma_{*}$ = 310 km s$^{-1}$, by comparing spectral lines sensitive to low-mass stars to the corresponding index strengths in the \cite{imf-conroy-3} SSP models. The mismatch value corresponding to a Chabrier IMF, log($\alpha_{\rm IMF}$) = -0.22, is adopted for all stellar particles with $\sigma_{*}$ less than 150 km s$^{-1}$.

\begin{figure}
  \includegraphics[width=1.\columnwidth]{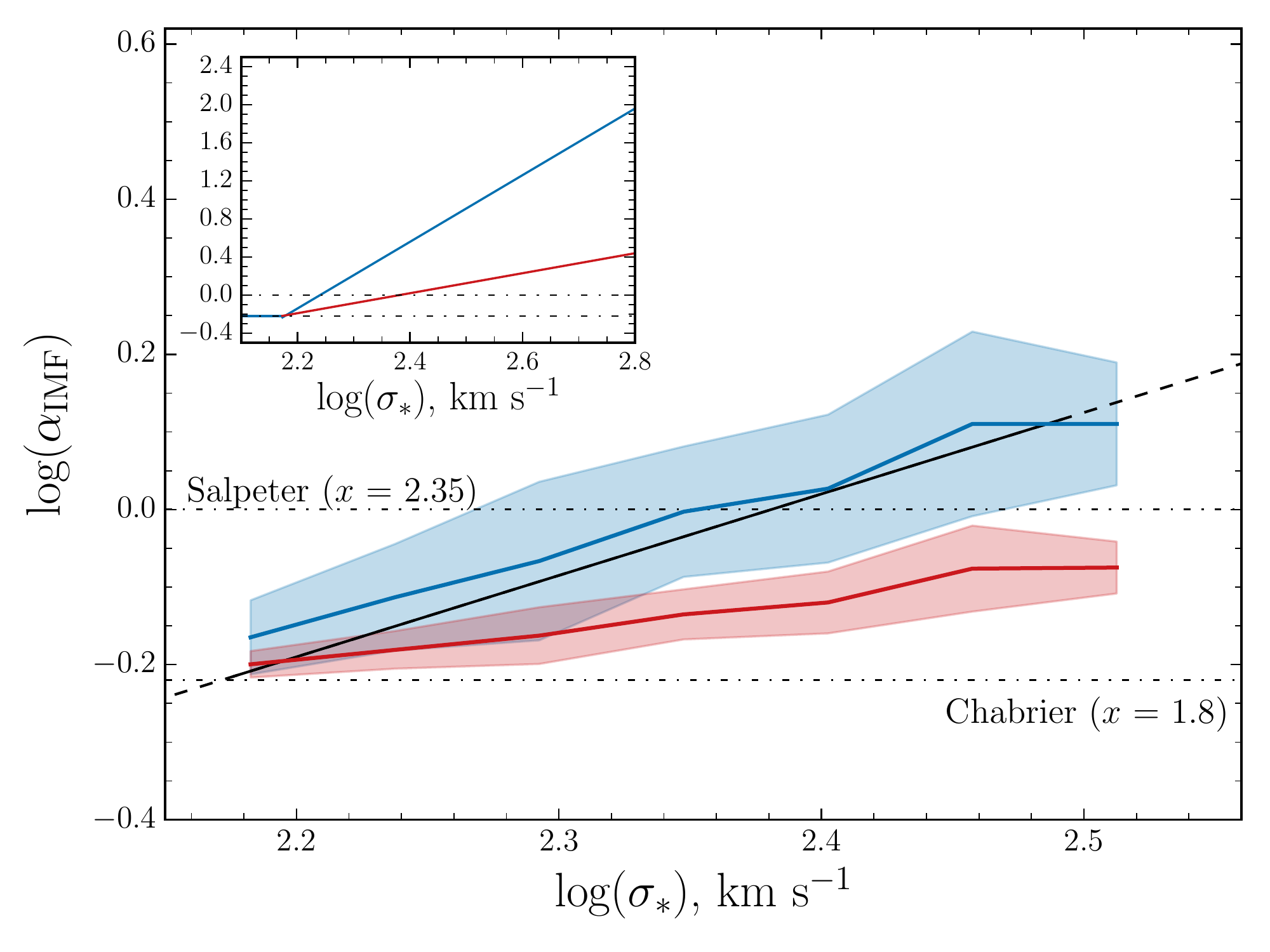}
      \caption{Main panel: IMF mismatch parameter, log($\alpha_{\rm IMF}$), as a function of $z = 0$ global stellar velocity dispersion, $\sigma_{*}$. Inset: the log($\alpha_{\rm IMF}$)-$\sigma_{*}$ relations used as input physical laws that produce the curves in the main panel. The solid black curve shows the observed relation from Spiniello et al. 2014 and the black dashed line shows an extrapolation of the relation. The resulting $z=0$ relations are always shallower than the input relations. Red: input relation as in Spiniello et al. 2014, blue: an input relation that is 3.5$\times$ steeper.
\bigskip }
    \label{fig:global_imf}
\end{figure}

\begin{figure*}
  \includegraphics[width=2.1\columnwidth]{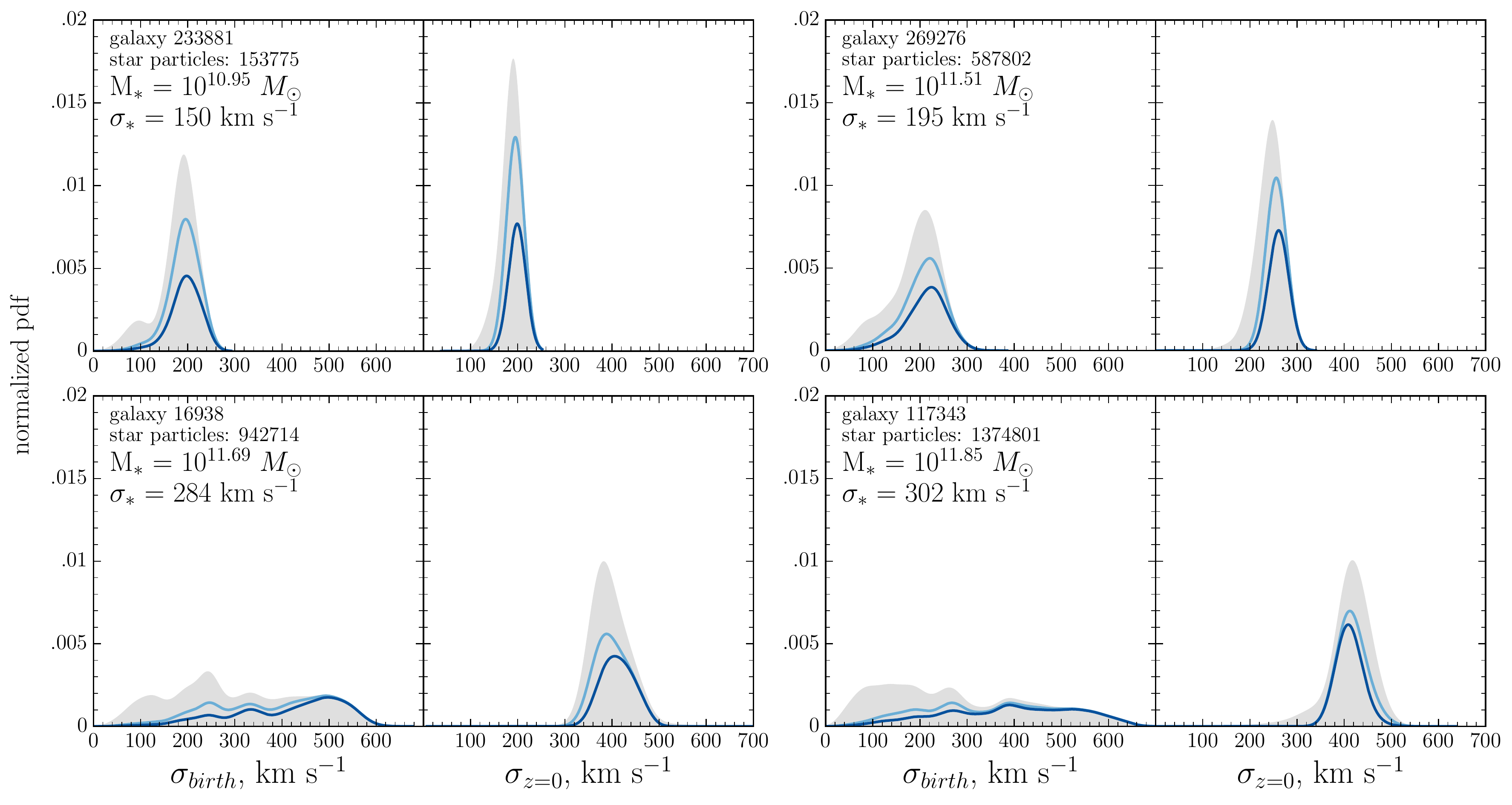}
    \caption{Velocity dispersion distributions of the stars in four Illustris galaxies both today (right panels) and at birth, namely in the first snapshot they appear in (left panels). Each pair of plots represents one galaxy, increasing in $z=0$ stellar mass from top-left to bottom-right. The grey curves correspond to all stars belonging to the $z=0$ galaxy, while the light (dark) blue curve corresponds to just the stellar particles within one (half) projected stellar half-mass radius $R^{p}_{1/2}$ (.5$R^{p}_{1/2}$) from the center of the $z=0$ galaxy. Evidently, the velocity dispersions of stars can change dramatically between their birth and $z=0$, especially in massive galaxies, which present very broad distributions of $\sigma_{\rm birth}$.
\bigskip }
    \label{fig:sigma_distr}
\end{figure*}

The red curve in Figure \ref{fig:global_imf} shows the resulting overall log($\alpha_{\rm IMF}$)-$\sigma_{*}$ relation for the $z=0$ Illustris galaxy sample. This relation was constructed based on the innermost parts of each galaxy, using only the stellar particles residing within 0.5$R^{p}_{1/2}$ from the center of the galaxy, to approximately match \cite{imf-spiniello}. The red curve in the inset shows the input relation used to construct log($\alpha_{\rm IMF}$), which is the same as the observed \cite{imf-spiniello} relation. The latter is repeated in the main panel as the black curve, to guide the eye. As evident in Figure \ref{fig:global_imf}, the \cite{imf-spiniello} relation applied at the time of stellar birth is not conserved through the assembly history of massive galaxies. The overall log($\alpha_{\rm IMF}$)-$\sigma_{*}$ relation is $\sim$2.5$\times$ too shallow compared to the observed log($\alpha_{\rm IMF}$)-$\sigma_{*}$ relation. While the observed overall log($\alpha_{\rm IMF}$) is reproduced for the lowest velocity dispersion galaxies in the sample, the constructed log($\alpha_{\rm IMF}$) of the higher velocity dispersion galaxies becomes increasingly too low. To reproduce the observed log($\alpha_{\rm IMF}$)-$\sigma_{*}$ relation, we construct input relations steeper than Equation \ref{eq:spiniello}. By minimizing the residuals between the resulting Illustris output relation and the \cite{imf-spiniello} relation, we determine input relation that produces the best-fit. The blue curves in Figure \ref{fig:global_imf} show that with an input relation 3.5$\times$ steeper than the observed relation (shown in the inset), the resulting overall log($\alpha_{\rm IMF}$)-$\sigma_{*}$ relation (shown in the main panel) \textit{is} able to reproduce both the slope and normalization of the observed trend within the reported errors.

This result suggests that the observed IMF variations with $z=0$ galactic velocity dispersion cannot be a correlation that exists in the galaxies in which the stars are actually born and where the IMF is set. This observed relation is hence emergent rather than fundamental. It is a manifestation of the complexities of $\Lambda$CDM galaxy formation through hierarchical assembly, where massive galaxies are composed of stellar populations that form in a plethora of progenitor galaxies with varied and evolving properties such as velocity dispersion. Since galaxies with high $\sigma_{*}$ contain stars that were formed inside galaxies with low $\sigma_{*}$ and hence have relatively bottom-light IMFs, it is necessary that stars forming in-situ in galaxies with high $\sigma_{*}$ have extremely bottom-heavy IMFs in order to combine together in their $z=0$ host galaxies and produce the observed relation.

In this section, we have used the same quantity on the horizontal axes of both the inset and the main panel, namely the assumed physical driver was the same as the independent variable of the $z=0$ relation. We have shown that the input relation is not preserved through galaxy assembly. For the remainder of this paper, we focus on connecting the global properties of $z=0$ galaxies to physical properties that are more closely associated with star-formation. We examine the local velocity dispersion and metallicity of the stellar particles, as well global quantities, star-formation rate and gas velocity dispersion, which have been suggested as drivers of IMF variations on the star-formation scale.

\subsection{Local velocity dispersion}
\label{sec:dmsigma}

Still motivated by the observed IMF trends with velocity dispersion, but aiming for a more physically relevant IMF driver, we construct the overall log($\alpha_{\rm IMF}$) of each Illustris galaxy based on the local velocity dispersion around each individual stellar particle at the time of formation. In particular, we use the one-dimensional velocity dispersion of the 64$\pm$1 dark matter particles nearest to each stellar particle in the earliest snapshot where it exists. This quantity, denoted as $\sigma_{\rm birth}$, probes the local gravitational potential at the time of stellar birth. 

Figure \ref{fig:sigma_distr} includes four example velocity dispersion distributions, comparing the local velocity dispersions the stars have in their $z=0$ host galaxy (right panels) to the velocity dispersions those same stars had at their individual formation times (left panels). We also distinguish between the distributions of local velocity dispersions for stellar particles enclosed within different radii with respect to the center of the $z=0$ galaxy, which is defined as the position of the most bound particle belonging to the galaxy. Figure \ref{fig:kde_grid} shows four additional $\sigma_{\rm birth}$ distributions for comparison to the various other stellar birth properties that will be discussed in following sections.

\begin{figure*}
  \includegraphics[width=2.\columnwidth]{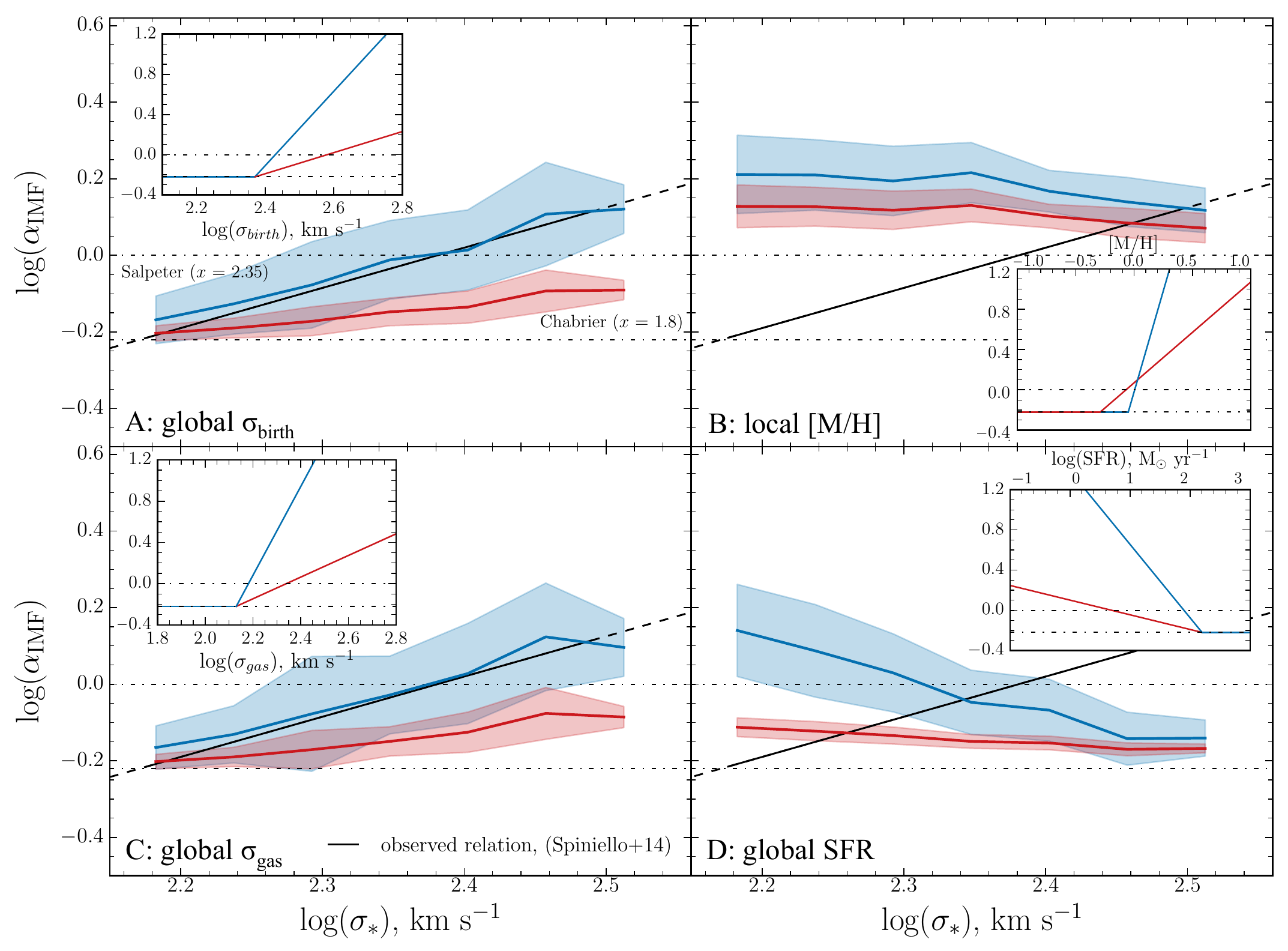}
      \caption{Relations between the IMF mismatch parameter log($\alpha_{\rm IMF}$) and various galaxy properties. The insets show the input relations used to construct log($\alpha_{\rm IMF}$), in each panel using a different physical quantity at star-formation time (\textbf{Panel A}: local velocity dispersion $\sigma_{\rm birth}$, \textbf{Panel B}: local metallicity [M/H], \textbf{Panel C}: global star-forming gas velocity dispersion $\sigma_{\rm gas}$, \textbf{Panel D}: global star-formation rate SFR). The main panels show the resulting constructed relations between log($\alpha_{\rm IMF}$) within 0.5$R^{p}_{1/2}$ and $z = 0$ global stellar velocity dispersion, $\sigma_{*}$. The red curves correspond to shallower input relations while the blue curves to steeper ones. In each panel the solid black line shows the observed relation from Spiniello et al. 2014 and the black dashed line shows an extrapolation of the relation. The dot-dashed lines at log($\alpha_{\rm IMF}$) = 0 and log($\alpha_{\rm IMF}$) = -0.22 indicate the Salpeter and Chabrier IMF mismatch parameters respectively. All input relations and fits to the output relations are listed in Table A1. \textit{Note:} In Panel D, the blue log($\alpha_{\rm IMF}$)-$\sigma_{*}$ relation has been shifted by -0.1 dex.
\smallskip }

   \label{fig:imf}
\end{figure*}

As evident in Figure \ref{fig:sigma_distr} the velocity dispersions that the stellar particles have in their host galaxy at $z = 0$ can be substantially different from the velocity dispersions they had at their time of birth.  Qualitatively, the $\sigma_{z=0}$ distributions for galaxies of all masses in our selected sample are singly peaked and best described as a gaussian or a gaussian with a low velocity dispersion tail.  In most cases the low velocity dispersion tail is largely built up by stellar particles residing outside the projected stellar half-mass radius of the galaxies. Unlike the $z = 0$ distributions, the birth velocity dispersion distributions are quite varied. Most lower mass galaxies in our sample ($M_{*}$  $\lessapprox$ 10$^{11}$ $M_{\odot}$) have near singly peaked birth distributions, spread out over a broader range of velocity dispersions than their $z = 0$ distributions. Some of these lower mass galaxies end up with a higher mean velocity dispersion at $z = 0$ than a birth, and others vice versa. Gas inflows, outflows, internal dynamical processes, as well as mergers, are all expected to play a role in shifting the velocity dispersion of a galaxy over time either to lower or higher values. 

Generally, the range of birth velocity dispersions becomes larger for galaxies with a higher $z = 0$ stellar mass. In particular, the birth velocity dispersion distributions of $M_{*}$  $\gtrapprox$ 10$^{11}$ $M_{\odot}$ galaxies in our sample are usually multi-peaked and spread across a large range of velocity dispersions. This is reflective of the rich merger histories of these high mass galaxies, with their stellar particles being formed in numerous progenitor galaxies with varying masses and velocity dispersions. For example, the $M_{*}$ = 10$^{11.69}$ $M_{\odot}$ galaxy shown in the bottom left panel of Figure \ref{fig:sigma_distr} underwent 4 major mergers ($\mu$ \textgreater ~1/4), 7 minor mergers (1/4 \textgreater ~$\mu$ \textgreater ~1/10), and 529 very minor mergers ($\mu$ \textless ~1/10) throughout its history. On the other hand, the lower mass galaxy shown in the top left panel, with $M_{*}$ = 10$^{10.95}$ $M_{\odot}$, only underwent 3 major mergers ($\mu$ \textgreater ~1/4), no minor mergers (1/4 \textgreater ~$\mu$ \textgreater ~1/10), and 96 very minor mergers ($\mu$ \textless ~1/10) throughout its history.

Radial trends are also present in the birth velocity dispersion distributions of massive galaxies, with the stellar particles closer to the center of each galaxy having, on average, higher birth velocity dispersions. This is due to the spatial distribution of stellar particles inside galaxies set up by mergers. As shown in \cite{r-gomez}, higher mass galaxies in Illustris consist of a larger fraction of stellar particles formed ex-situ, i.e.~not on the main progenitor branch. Galaxies with stellar masses greater than 10$^{12}$ M$_{\odot}$ can have up to 80\% of their stellar particles formed ex-situ and later accreted onto the main galaxy via merging. \cite{r-gomez} finds that stellar particles formed in-situ tend to reside in the innermost regions of galaxies whereas stars formed ex-situ tend to lie in the outer regions at larger galactocentric distances.

With the birth velocity dispersion and mass of each stellar particle belonging to a galaxy, we construct the IMF mismatch parameter according to the prescription in Section \ref{sec:imf}. To start, we shift the observed relation presented by \cite{imf-spiniello} towards higher velocity dispersions,

\begin{equation} \label{eq:shifted}
\mathrm{log}(\alpha_{\rm IMF}) = 1.05\mathrm{log}(\sigma_{\rm birth}) - 2.71,
\end{equation}

\noindent so that the minimum Chabrier IMF value of $\alpha_{\rm IMF}$ = 0.6 is adopted for all stellar particles with $\sigma_{\rm birth}$ less than 235 km s$^{-1}$. This shift in the input relation is applied because the $\sigma_{\rm birth}$ distributions, on average, cover higher values than the $\sigma_{*}$ distributions. A shift in the relation is needed to place the log($\alpha_{\rm IMF}$) of low velocity dispersion galaxies on the observed log($\alpha_{\rm IMF}$)-$\sigma_{*}$ relation.\footnote{Using the original \cite{imf-spiniello} relation as input in this case is still unable to reproduce the slope of the observed relation.}. 

Panel A of Figure \ref{fig:imf} shows the resulting log($\alpha_{\rm IMF}$)-$\sigma_{*}$ relation for $z = 0$ Illustris galaxies constructed based on $\sigma_{\rm birth}$. We show only the relations constructed using the stellar particles residing within 0.5$R^{p}_{1/2}$ of each $z = 0$ galaxy, as observations of IMF variations are mainly constructed using the innermost regions of galaxies, and the same holds for the other panels in Figure \ref{fig:imf} as well. For the output relations constructed using all the stellar particles belonging to each galaxy, the reader is referred to Table A1. The inset figure in each panel of Figure \ref{fig:imf} shows the input relations we used to construct log($\alpha_{\rm IMF}$) based on the indicated stellar particle property. The red curve in Panel A shows the log($\alpha_{\rm IMF}$)-$\sigma_{*}$ trend resulting from using Equation \ref{eq:shifted} as the prescribed IMF relation applied to the birth velocity dispersions of each galaxy's stellar particles. As evident in the figure, the output log($\alpha_{\rm IMF}$)-$\sigma_{*}$ is $\sim$2.8$\times$ too shallow compared to the observed log($\alpha_{\rm IMF}$)-$\sigma_{*}$ relation (black curve). This can be understood in terms of the birth velocity dispersion distributions shown in Figure \ref{fig:sigma_distr}. Using Equation \ref{eq:shifted} as the input relation, there is not enough of a differentiation created between the overall log($\alpha_{\rm IMF}$) of galaxies of different global $\sigma_{*}$ values. For example, the $\sigma_{*}$ = 150 km s$^{-1}$ and $\sigma_{*}$ = 195 km s$^{-1}$ galaxies shown in Figure \ref{fig:sigma_distr} have birth velocity dispersion distributions that cover a similar range of values. So the log($\alpha_{\rm IMF}$) difference between the two galaxies, where the $\sigma_{*}$ = 150 km s$^{-1}$ galaxy is found to have an overall log($\alpha_{\rm IMF}$) = -0.219 and the $\sigma_{*}$ = 195 km s$^{-1}$ galaxy is found to have log($\alpha_{\rm IMF}$) = -0.210, is too small compared to the observed difference of $\Delta$log($\alpha_{\rm IMF}$) = .125.  

To reproduce the observed log($\alpha_{\rm IMF}$)-$\sigma_{*}$ relation, we construct various input log($\alpha_{\rm IMF}$)-$\sigma_{\rm birth}$ relations with steeper slopes. As before, we minimize the residuals between the resulting Illustris output relations and the \cite{imf-spiniello} relation, and select the input relation that produces the best-fit. The blue curve in Panel A of Figure \ref{fig:imf} shows the resulting log($\alpha_{\rm IMF}$)-$\sigma_{*}$ relation constructed using an input relation that is $\sim$3.5$\times$ steeper than Equation \ref{eq:shifted} and a minimum Chabrier $\alpha_{\rm IMF}$ applied to stellar particles with $\sigma_{\rm birth}$ \textless ~235 km s$^{-1}$. As seen in Panel A of Figure \ref{fig:imf}, this steeper input relation is able to reproduce both the slope and normalization of the observed trend with global velocity dispersion. The increase in the slope of the input relation creates more of a differentiation between galaxies of different $\sigma_*$. For example, the $\sigma_{*}$ = 150 km s$^{-1}$ galaxy in Figure \ref{fig:sigma_distr} is found to have a similar overall log($\alpha_{\rm IMF}$) as before (log($\alpha_{\rm IMF}$) = -0.217), but the $\sigma_{*}$ = 195 km s$^{-1}$ overall log($\alpha_{\rm IMF}$) increased by $\Delta$log($\alpha_{\rm IMF}$) = .025, placing it closer to the observed log($\alpha_{\rm IMF}$)-$\sigma_{*}$ relation.

\begin{figure*}
  \includegraphics[width=2.\columnwidth]{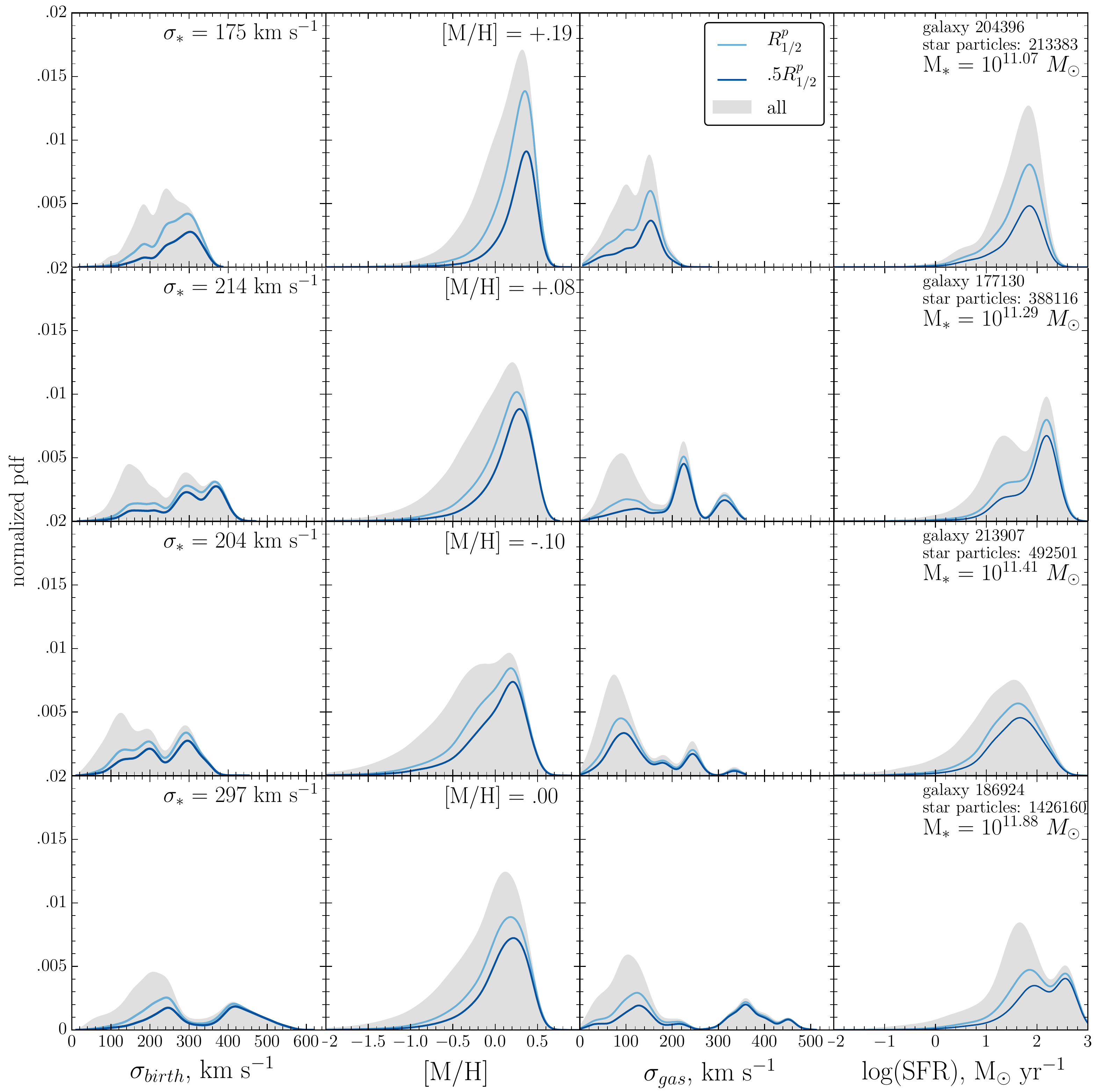}
      \caption{Birth properties of stellar particles belonging to four $z = 0$ galaxies, showing how various stellar birth property distributions vary with galaxy mass. Each row shows the properties of one galaxy, increasing in stellar mass from top to bottom where \textbf{Column 1}: local, birth velocity dispersion ($\sigma_{\rm birth}$), \textbf{Column 2}: birth metallicity ([M/H]), \textbf{Column 3}: gas velocity dispersion ($\sigma_{\rm gas}$), \textbf{Column 4}: star-formation rate.The grey distributions show the properties of all the stellar particles belonging to the $z = 0$ galaxy while the the light blue and dark blue distributions show the properties of the stellar particles within one and one-half the projected stellar half-mass radius, respectively.
\smallskip }
    \label{fig:kde_grid}
\end{figure*}

In addition to increasing the slope of the output log($\alpha_{\rm IMF}$)-$\sigma_{*}$ relation, a steeper input relation acts to increase the scatter of the Illustris log($\alpha_{\rm IMF}$)-$\sigma_{*}$ relation. As seen in Panel A of Figure \ref{fig:imf}, the scatter of the blue relation is nearly 2$\times$ the scatter of the red relation. The increase in scatter with the steeper input relation can also be understood with an example from Figure \ref{fig:sigma_distr}. Although the 10$^{11.69}$ $M_{\odot}$ and 10$^{11.85}$ $M_{\odot}$ galaxies shown in Figure \ref{fig:sigma_distr} have similar masses and velocity dispersions, their birth velocity dispersion distributions are quite different. Using Equation \ref{eq:shifted} as the input relation, the difference between their overall log($\alpha_{\rm IMF}$) values is $\Delta$log($\alpha_{\rm IMF}$) = .025. But, using the steeper input relation the difference becomes $\Delta$log($\alpha_{\rm IMF}$) = .10. While a steeper input relation is able to create more of an overall differentiation between galaxies of different $\sigma_{*}$ values, it also creates a larger scatter among galaxies of similar $\sigma_{*}$ but different formation histories.

%%%%%%%%%%%%%%%%%%%% METALLICITY %%%%%%%%%%%%%%%%%%

\subsection{Local metallicity}
\label{sec:metallicity}

The second physical quantity of star-formation we examine is metallicity. Aside from velocity dispersion, observational studies also report that the IMF scales with galaxy metallicity. While both velocity dispersion and metallicity correlate with mass, the quantity that is more fundamentally associated with IMF variations is still unclear.

A metallicity-IMF correlation can be easily imagined through a reversed causal relationship. Simply put, the IMF is expected to influence the metallicity because the number of high to low mass stars will directly affect the chemical evolution of a galaxy. The more top-heavy the IMF, the more metals are injected into the ISM, which increases the metallicity of the stars born in subsequent star formation bursts. 

So one might naively expect the overall metallicity of a galaxy by $z = 0$ to be higher with a more top-heavy its IMF. This scenario is however in tension with observational IMF-metallicity relations, which infer a more {\it bottom}-heavy IMF for the most metal-rich galaxies \citep[e.g.][]{imf-conroy-1, imf-martin-1}. One way this tension may be reconciled is by invoking a time-dependent IMF: earlier star formation follows a flatter IMF to build up the metallicity of the ISM and star formation occurring later follows a bottom-heavy IMF to build up the population of low-mass stars \citep{imf-martin-3, imf-weidner, imf-narayanan}. Particularly, \cite{imf-weidner} propose that the ISM, enriched by episodes of high star-formation with a flat IMF, is exceptionally turbulent leading to increased fragmentation on lower mass scales. Therefore, proceeding star-formation occurs with a steeper IMF slope. In this scenario, a higher metallicity environment at the time of stellar birth is expected to correspond to a more bottom-heavy IMF. 

Our current empirical approach for constructing log($\alpha_{\rm IMF}$) does not take into account how the IMF may influence metallicity. Instead, motivated by \cite{imf-weidner}, we assign an IMF based on the local metallicity of each stellar particle at the time of stellar birth with the idea that stellar particles born into higher metallicity environments form with a steeper IMF slope. Thus, we construct the overall log($\alpha_{\rm IMF}$) of each Illustris galaxy based on the metallicity of each stellar particle at the time of formation. As discussed in Section \ref{sec:illustris}, the total mass in metals of a stellar particle in Illustris is inherited from the parent gas cell at the time of star formation. The ratio of the total mass in metals heavier than helium to the total mass of the stellar particles at the time of birth, $Z$, is output for each star in the snapshot files.  For the stellar particles in our sample, we convert the metallicity mass fraction to the metal abundance [M/H] by assuming each stellar particle to have a primordial hydrogen mass fraction of $X$ = 0.76 and scaling to solar units using $Z_{\odot} = 0.02$ and $X_{\odot} = 0.70$.  

The second column of Figure \ref{fig:kde_grid} shows the distribution of stellar particle metallicities for four Illustris galaxies, increasing in $z = 0$ stellar mass from top to bottom. The light grey histogram shows the distribution of metallicities of all stellar particles belonging to the galaxy while the light and dark blue distributions show the metallicities of the stellar particles within $R^{p}_{1/2}$ and 0.5$R^{p}_{1/2}$, respectively. The distributions are similarly shaped, mostly described as a gaussian with a significant low metallicity tail. Similar to the velocity dispersion distributions in Figure \ref{fig:sigma_distr}, the stellar particles residing closer to the center of the galaxy have a higher average metallicity than the stellar particles residing in the outer reaches of the galaxy. The global $z = 0$ metallicity of each galaxy, shown in the upper right of each panel, is calculated by taking the mass-weighted average of the metal abundance of the stellar particles within 0.5$R^{p}_{1/2}$ of each galaxy.

\begin{figure}
  \includegraphics[width=1.\columnwidth]{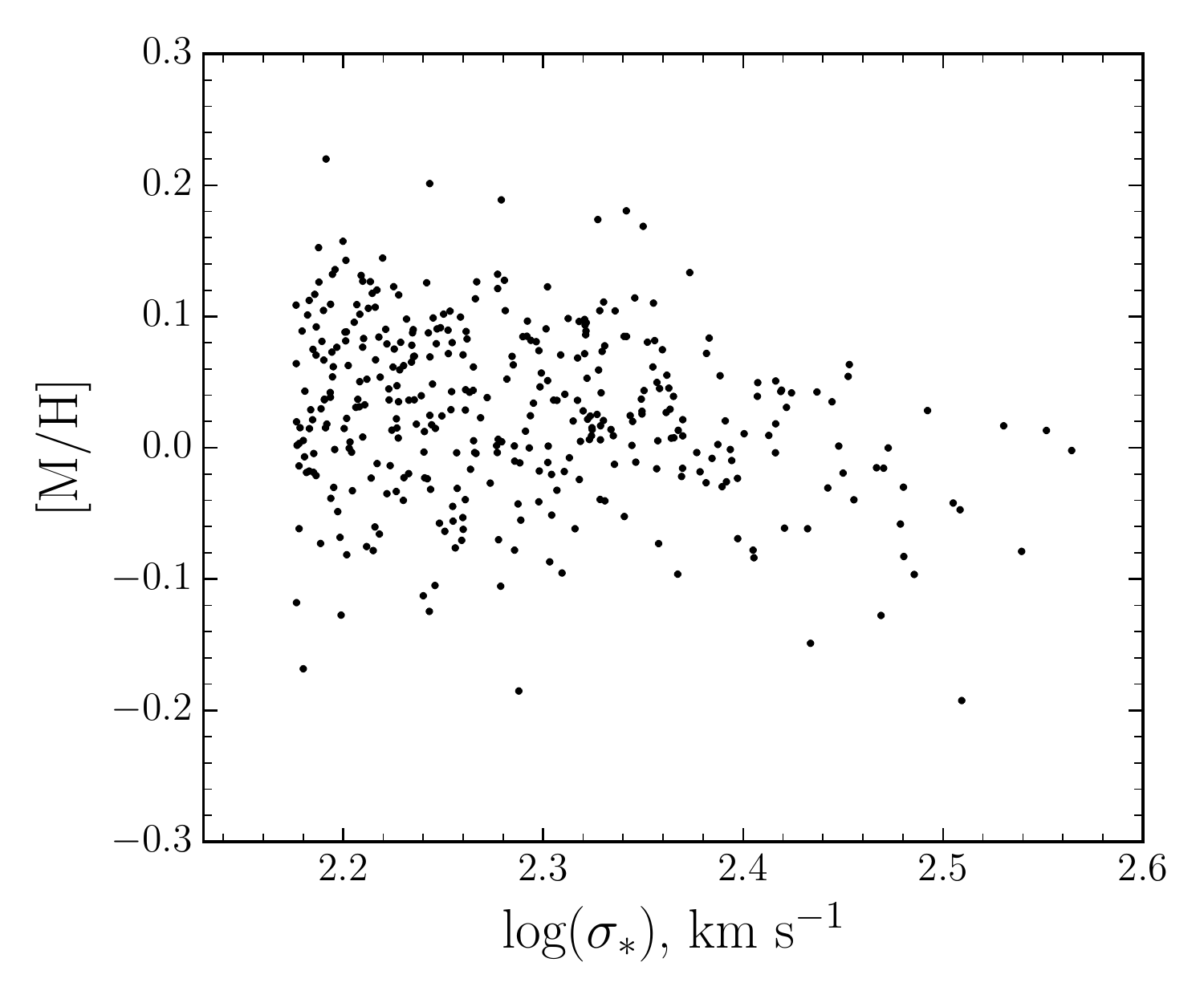}
    \caption{Global metallicity, [M/H], as a function of $z = 0$ stellar velocity dispersion, $\sigma_{*}$ for the 371 selected Illustris-1 galaxies.}
    \label{fig:metals}
\end{figure}

With the birth metallicities and mass of each stellar particle belonging to a galaxy, we construct the IMF mismatch parameter as before. As a starting point, we are inspired by \cite{imf-martin-2} to construct an input log($\alpha_{\rm IMF}$)-[M/H] relation that is defined to have a Chabrier $\alpha_{\rm IMF}$ value at [M/H] = -0.29 and a Salpeter $\alpha$ value at [M/H] = -0.07. This results in the relation log($\alpha_{\rm IMF}$) = [M/H] + 0.07, where stellar particles with metallicities less than -0.29 are assigned $\alpha_{\rm IMF}$ = 0.6. Panel B of Figure \ref{fig:imf} shows the resulting Illustris log($\alpha_{\rm IMF}$)-$\sigma_{*}$ relation (red curve) constructed based on the individual [M/H] values that each stellar particle belonging to a galaxy are formed with. Again, we only show the log($\alpha_{\rm IMF}$)-$\sigma_{*}$ relations constructed using only the stellar particles residing within 0.5$R^{p}_{1/2}$ of each $z = 0$ galaxy. Evidently, our initial input relation is unable to reproduce the observed log($\alpha_{\rm IMF}$)-$\sigma_{*}$ relation, with the slope of the output Illustris relation being nearly flat and in fact slightly negative, showing the opposite trend to the observed one.

To try to reproduce the observed relation, we construct an input log($\alpha_{\rm IMF}$)-[M/H] that is 4$\times$ as steep and defined to have a minimum Chabrier $\alpha$ at [M/H] = -0.05. The increase in the Chabrier minimum is in attempt to decrease the overall log($\alpha_{\rm IMF}$) values of lower velocity dispersion galaxies. The blue curve in Panel B of Figure \ref{fig:imf} shows the log($\alpha_{\rm IMF}$)-$\sigma_{*}$ relation constructed using the steeper log($\alpha_{\rm IMF}$)-[M/H] described. The result is that the normalization of the output relation shifts to higher log($\alpha_{\rm IMF}$) values, but its slope is still negative. Additionally, increasing the Chabrier minimum to [M/H] = -0.05 did not act to decrease the overall log($\alpha_{\rm IMF}$) of lower velocity dispersion galaxies, as increasing the Chabrier minimum in Section \ref{sec:dmsigma} was able to do. 

The inability to reproduce the observed IMF trend with $z = 0$ velocity dispersion using [M/H] as the physical driver of the IMF can be understood from the global [M/H]-$\sigma_{*}$ relation. As seen in Figure \ref{fig:metals}, this relation is almost flat for our sample comprised of massive $M_{*}$ \textgreater ~$\sim$10$^{10}$ $M_{\odot}$ galaxies\footnote{The flatness of the [M/H]-$\sigma_{*}$ relation is generally in agreement with observations of the stellar mass-metallicity relation, such as in \cite{r-gallazzi}, where the SDSS mass-metallicity becomes flat at high stellar masses with a scatter of $\sim$0.3 dex. Figure 5 is also in agreement with the observed velocity dispersion-mass relation over the appropriate velocity dispersion range \citep{imf-spolaor}.} In fact, where it is not {\it completely} flat, at $2.4<\log{\sigma_{*}}<2.6$, the output log($\alpha_{\rm IMF}$)-$\sigma_{*}$ also has some slope. The negative slope of the [M/H]-$\sigma_{*}$ relation could either be due to intracluster light contamination or recycling of low metallicity gas. However, even there, the mild slope of the [M/H]-$\sigma_{*}$ relation combined with its large scatter result in galaxies of similar velocity dispersions having a wide range of global [M/H] values. Since the global [M/H] of each galaxy is related to the distribution of the individual stellar particle [M/H] values (as seen in Figure \ref{fig:kde_grid}) \textit{and} the widths of the [M/H] distributions are broad compared to the galaxy-to-galaxy differences in global [M/H], a differential effect in overall log($\alpha_{\rm IMF}$) between galaxies of different velocity dispersions is not produced using an input relation based on [M/H].

%%%%%%%%%%%%%%%%%%%% SF GAS %%%%%%%%%%%%%%%%%%

\subsection{Global star-forming gas velocity dispersion}
\label{sec:sfgas}

As mentioned in Section \ref{sec:introduction}, a few analytical studies have focused on connecting the physics governing star formation to an environment dependent IMF. In particular, \cite{r-hopkins} developed an analytical formulation where star-forming disks with higher Mach numbers cause the low-mass turnover of the pre-stellar core mass function (CMF) to be shifted to lower masses leading to a more bottom-heavy CMF which implies a more bottom-heavy IMF. Physically, as discussed in \cite{r-hopkins}, a higher star-forming disk Mach number leads to larger density fluctuations which causes more fragmentation on smaller mass scales. 

Motivated by a Mach number dependent CMF, we construct the IMF of our Illustris galaxies using the one-dimensional, global star-forming gas velocity dispersion ($\sigma_{\rm gas}$) of the progenitor galaxies in which stellar particles are born. We do not take into account differences in sound speed, but simply use $\sigma_{\rm gas}$ as a proxy for Mach number \citep{imf-chabrier-3}. To calculate $\sigma_{\rm gas}$ we trace each stellar particle back to the progenitor galaxy in which it was born and consider only the gas cells in that galaxy with non-zero instantaneous star formation rates. We remove net rotation by calculating the total angular momentum vector of each galaxy's star-forming gas component and calculate $\sigma_{\rm gas}$ as the mass-weighted standard deviation of the cell velocities parallel to that angular momentum vector.

The third column of Figure \ref{fig:kde_grid} shows the $\sigma_{\rm gas}$ distributions for four galaxies in our sample. The birth $\sigma_{\rm gas}$ distributions are similarly multi-peak and spread across a broad range of values as the $\sigma_{\rm birth}$ distributions shown in the first column of Figure \ref{fig:kde_grid}.  However, there are a few qualitative differences between the two distributions. First, the $\sigma_{\rm gas}$ distributions are less continuous than the corresponding $\sigma_{\rm birth}$ distributions, reflective of the fact that multiple stellar particles are often born in the same progenitor galaxy and therefore have the same $\sigma_{\rm gas}$ value. The $\sigma_{\rm gas}$ distributions are also shifted to lower velocity dispersion values compared to their $\sigma_{\rm birth}$ counterparts due to the removal of rotation. But similar to the $\sigma_{\rm birth}$ distributions, there is a radial trend in $\sigma_{\rm gas}$ especially for higher mass galaxies, with a larger fraction of stars born in high $\sigma_{\rm gas}$ galaxies residing closer to the center of their $z = 0$ host galaxy.

We construct the overall log($\alpha_{\rm IMF}$) of each galaxy based on the $\sigma_{\rm gas}$ distributions. We first construct a log($\alpha_{\rm IMF}$)-$\sigma_{\rm gas}$ relation inspired by Equation \ref{eq:shifted}, but shift the Chabrier minimum of the relation to occur at $\sigma_{\rm gas}$ = 135 km s$^{-1}$. This results in the relation log($\alpha_{\rm IMF}$) = 1.05log($\sigma_{\rm gas}$) - 2.46, where stellar particles with $\sigma_{\rm gas}$ less than 135 km s$^{-1}$ are assigned $\alpha_{\rm IMF}$ = 0.6. Panel C of Figure \ref{fig:imf} shows the resulting Illustris log($\alpha_{\rm IMF}$)-$\sigma_{*}$ relation (red curve) constructed based on the star-forming gas velocity dispersion of the progenitor galaxy that each stellar particle belonging to a $z = 0$ was formed in, showing just the relation constructed using the stellar particles residing within 0.5$R^{p}_{1/2}$. As with $\sigma_{\rm birth}$, the initial input relation is unable to reproduce the observed log($\alpha_{\rm IMF}$)-$\sigma_{*}$ relation, with the slope of the Illustris log($\alpha_{\rm IMF}$)-$\sigma_{*}$  $\sim$2.6$\times$ shallower than the observed relation.

To reproduce the observed relation, we construct a steeper log($\alpha_{\rm IMF}$)-$\sigma_{\rm gas}$ relation. The blue curve in Panel C of Figure \ref{fig:imf} shows the log($\alpha_{\rm IMF}$)-$\sigma_{*}$ relation constructed using a log($\alpha_{\rm IMF}$)-$\sigma_{\rm gas}$ input relation that is 4.1$\times$ steeper than the \cite{imf-spiniello} relation. This steeper input relation does, within the reported uncertainty, reproduce both the slope and normalization of the observed trend with global velocity dispersion $\sigma_{*}$. As with $\sigma_{\rm birth}$ as a physical driver, the increase in the slope of the input relation creates more of a differentiation between galaxies of different $\sigma_{*}$, which allows the observed relation to be reproduced. Also similar to the Illustris log($\alpha_{\rm IMF}$)-$\sigma_{*}$ relation constructed based on $\sigma_{\rm birth}$, increasing the slope of the input relation results in a larger scatter in the output relation. This is because galaxies of similar $z = 0$ velocity dispersions can have a range of $\sigma_{\rm gas}$ distributions.

\subsection{Global star-formation rate}
\label{sec:sfr}

The last physical quantity we consider in constructing the overall log($\alpha_{\rm IMF}$) of each Illustris galaxy is star-formation rate. Observationally, studies focusing on constraining the high mass end of the IMF suggest that the IMF correlates with galaxy SFR. For example, \cite{imf-gunawardhana} find that for a range of galaxies at $z$ \textless ~0.35 with SFRs covering 10$^{-3}$ to 100 $M_{\odot}$ yr$^{-1}$, the most quiescent galaxies are best described by steeper IMF slopes ($x$$\sim$2.4), whereas highly star-forming galaxies exhibit shallower IMFs ($x$$\sim$1.8). They translate their IMF-SFR to a relation between IMF and SFR surface density, finding that galaxies with higher SFR densities prefer flatter IMF slopes. This result is consistent within the context of IGIMF theory \citep{imf-weidner-2}, which connects the global SFR of a galaxy to the formation of stars within individual molecular clouds throughout the galaxy, where galaxies with higher SFR are expected to have a top-heavy galaxy-wide IMF.

On the other hand, \cite{imf-conroy-1} find their strongest IMF trend to be with [Mg/Fe], where galaxies with greater Mg enhancement have more bottom-heavy IMFs. These galaxies with increased Mg abundances are interpreted as having shorter star-formation timescales. Based on the inferred star-formation time scales of massive galaxies with enhanced Mg abundances, \cite{imf-conroy-1} infer that galaxies with high SFR surface densities are described by a more bottom-heavy IMF. Physically, as pointed out by \cite{imf-conroy-1}, in the context of the \cite{r-hopkins} analytical theory for IMF variations, high SFR surface densities promotes turbulence which leads to a more bottom-heavy IMF. In regards to observations reporting that higher SFRs correspond to shallower IMFs and to IGIMF theory which predicts the same, it is suggested that the high $M_{*}/L$ ratio inferred for these massive elliptical galaxies is at least in part due to an excess of high-mass stellar remnants and not only an excess of low-mass stars. In this scenario, high SFR starbursts induced by mergers form with a top-heavy IMF, and it is the remnants of these high mass stars that produce an excess of mass as measured by $z = 0$.  

For each stellar particle that comprises a $z = 0$ galaxy, we record the instantaneous star-formation rate of the progenitor galaxy in which the stellar particle is formed. The fourth column of Figure \ref{fig:kde_grid} shows the distribution of birth SFRs for the stellar particles comprising four galaxies of various masses. As seen in the figure, the birth SFR distribution becomes multi-peaked and/or broader for galaxies with higher stellar mass, and includes more stellar particles with higher birth SFRs. For lower mass galaxies, where a majority of their stellar populations are formed in-situ, the shape of the birth SFR distribution can be understood as the evolution of the star-formation rate of the main progenitor branch. The low SFR tails of these distributions correspond to the formation of stellar particles before and after the period of peak star-formation where most of the stellar mass is formed. Additionally, for these lower mass galaxies, there is little difference in the birth SFR distributions of all the stellar particles versus just the stellar particles residing within 0.5$R^{p}_{1/2}$ of each galaxy. 

The higher mass galaxies shown in the figure ($M_{*}$ \textgreater ~10$^{11.2}$ $M_{\odot}$) have stellar particles that, on average, formed in progenitor galaxies with higher SFRs and also cover a broader range of SFRs. The stellar particles formed in progenitor galaxies with SFRs $\sim$ 100 $M_{\odot}$ yr$^{-1}$ likely formed during merger-induced nuclear starbursts, whereas the stellar particles making up the lower SFR part of the distributions were formed either before or after the peak star-formation period of these merger events, or in lower SFR galaxies that are later accreted onto the main progenitor. The SFR for the 10$^{11.88}$ $M_{\odot}$ and 10$^{11.29}$ $M_{\odot}$ galaxies particularly show that stellar particles residing closer to the center of the $z = 0$ galaxy are formed in galaxies with higher SFRs than stellar particles residing in the outer edges of the galaxy. This is consistent with the $\sigma_{\rm birth}$ distribution of massive galaxies, where stellar particles closer to the center of a massive galaxy are formed in high velocity dispersion environments during periods of high SFR nuclear starbursts. 

We construct the log($\alpha_{\rm IMF}$)-$\sigma_{*}$ relation for our sample of Illustris galaxies based on the birth SFR distributions described above. First, we use a log($\alpha_{\rm IMF}$)-SFR relation that is defined to correspond to a Chabrier IMF at log(SFR) = 2.2 and a Salpeter IMF at log(SFR) = 0.7. This starting point is inspired by \cite{imf-gunawardhana} who for their sample of galaxies from GAMA find an IMF-SFR relation -x $\approx$ 0.36 log(SFR) - 2.6. Stellar particles born into galaxies with SFR = 0 $M_{\odot}$ yr$^{-1}$, where the SFR is an unresolved small value, are assigned zero light, i.e. $\alpha_{\rm IMF}$$^{-1}$ = 0. The red curve in Panel D of Figure \ref{fig:imf} shows the log($\alpha_{\rm IMF}$)-$\sigma_{*}$ relation resulting from this initial input relation. As seen in the figure, using SFR as the physical quantity of star-formation, where high SFR environments are expected to correspond to a shallower IMF, to build the Illustris log($\alpha_{\rm IMF}$)-$\sigma_{*}$ results in a trend opposite to that of the observations. Inputting a $\sim$4.7$\times$ steeper log($\alpha_{\rm IMF}$)-SFR relation, we are best able to reproduce the steepness of the observed log($\alpha_{\rm IMF}$)-$\sigma_{*}$ trend, but in the opposite direction (blue curve). This is because higher velocity dispersion galaxies in our sample have stellar particles that, on average, formed in progenitors with higher SFRs. This is compared to the stellar particles belonging to low velocity dispersion galaxies, which generally form in progenitors with lower SFRs.

\section{Results beyond the global \lowercase{$z$} = 0 trends}
\label{sec:results2}

\subsection{Radial trends}
\label{sec:radial}

As suggested by Figures \ref{fig:sigma_distr} and \ref{fig:kde_grid}, the physical quantities that we investigate display various amounts of radial variation depending on the mass of the galaxy. For the more massive galaxies in our sample, stellar particles residing closer to the centers of their $z = 0$ galaxy tend to have higher $\sigma_{\rm birth}$, [M/H], $\sigma_{\rm gas}$, or SFRs compared to the stellar particles residing in the outskirts of the galaxy. Such a radial trend is  weaker or non-existent for the lower mass galaxies we examine. This is reflective of lower mass galaxies being composed of a smaller fraction of stellar particles formed ex-situ compared to high mass galaxies. So in constructing the overall log($\alpha_{\rm IMF}$) using only stellar particles within 0.5$R^{p}_{1/2}$, the log($\alpha_{\rm IMF}$)-$\sigma_{*}$ relation is steeper than when constructing log($\alpha_{\rm IMF}$) based on all the stellar particles belonging to each galaxy. Table \ref{tab:relations} lists the comparison between the output log($\alpha_{\rm IMF}$)-$\sigma_{*}$ relations constructed using all stellar particles versus just the innermost stellar particles.

\begin{figure}
  \includegraphics[width=1.\columnwidth]{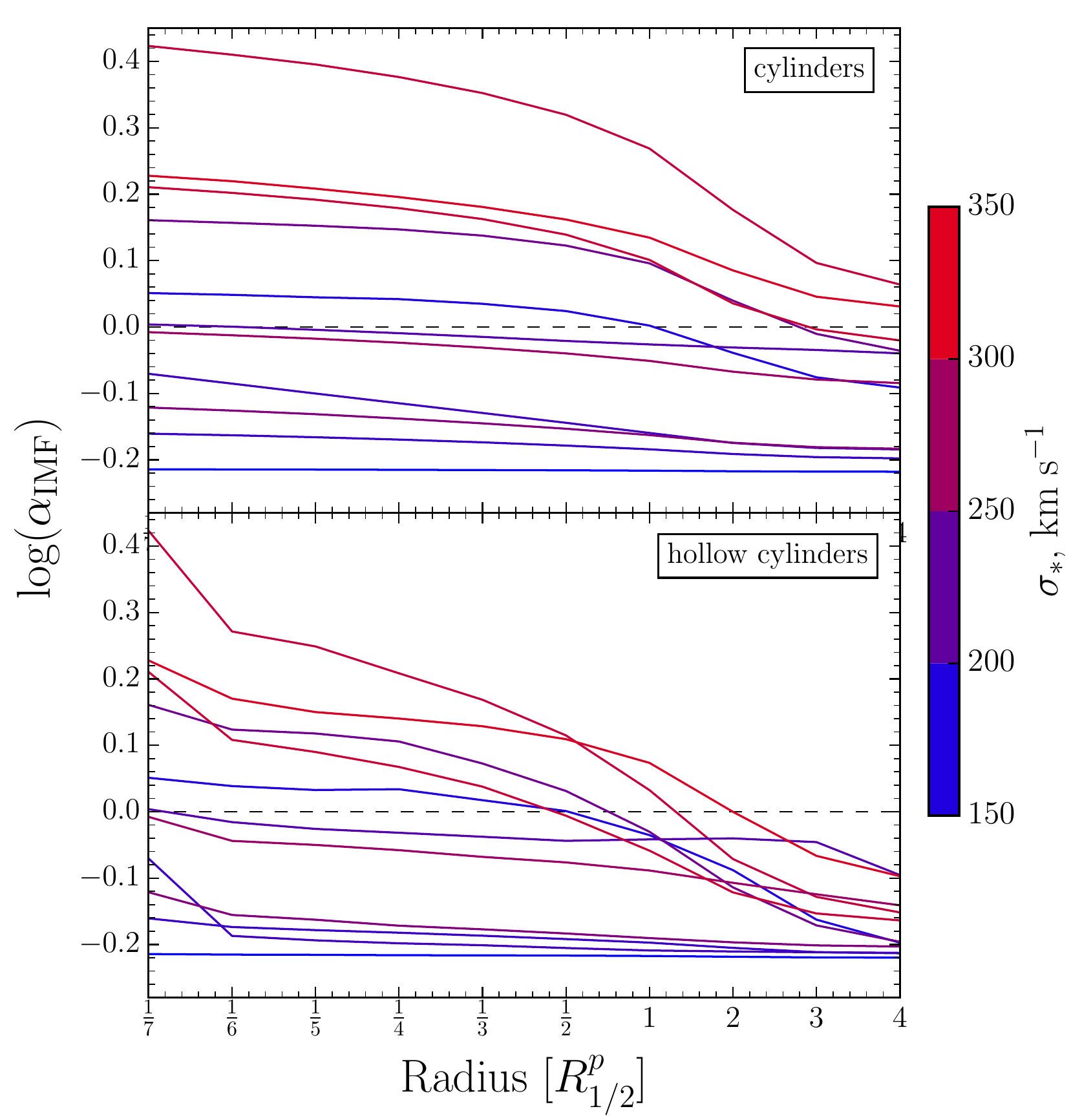}
    \caption{IMF slope as a function of galactocentric distance for 11 galaxies in our sample, using the steep log($\alpha_{\rm IMF}$)-$\sigma_{\rm birth}$ relation (see text). Each line represents a galaxy colored by its $z = 0$ velocity dispersion.  The top panel shows the IMF slope constructed in cylinders, including all stars within each radius and the bottom panel shows the IMF slope constructed in hollow cylinders, including the stellar particles between two consecutive radii. These constructed Illustris log($\alpha_{\rm IMF}$) radial gradients are in qualitative agreement with observations of IMF gradients for galaxies of both low and high velocity dispersion.
\medskip }
    \label{fig:radial_imf}
\end{figure}

Here we examine in more detail radial trends of log($\alpha_{\rm IMF}$) for a sample of galaxies with varying global velocity dispersions. To construct the overall IMF mismatch parameter we use the steep log($\alpha_{\rm IMF}$)-$\sigma_{\rm birth}$ relation, log($\alpha_{\rm IMF}$) = 3.7log($\sigma_{\rm birth}$) - 8.99, for which we were able to reproduce the observed log($\alpha_{\rm IMF}$)-$\sigma_{*}$ trend. Figurfe \ref{fig:radial_imf} shows log($\alpha_{\rm IMF}$) as a function of projected radius for 11 Illustris galaxies, starting at a radius of 1/7 the projected stellar half-mass radius, (1/7)$R^{p}_{1/2}$, out to radius of 4 times the projected stellar half-mass radius, 4$R^{p}_{1/2}$. The top panel shows log($\alpha_{\rm IMF}$) constructed in cylinders (i.e. spheres projected along the line of sight), including all stellar particles falling within the indicated radius. The bottom panel shows log($\alpha_{\rm IMF}$) constructed in hollow cylinders, where each indicated radius shows log($\alpha_{\rm IMF}$) constructed with just the stellar particles falling within two consecutive radii. The log($\alpha_{\rm IMF}$) value for (1/7)$R^{p}_{1/2}$ is constructed using stellar particles residing between (1/8)$R^{p}_{1/2}$ and (1/7)$R^{p}_{1/2}$.  

As evident in Figure \ref{fig:radial_imf}, the highest velocity dispersion galaxies ($\sigma$ $\sim$ 250 - 350 km s$^{-1}$) exhibit the greatest decrement in log($\alpha_{\rm IMF}$) towards larger radii, whereas log($\alpha_{\rm IMF}$) for lower velocity dispersion galaxies ($\sigma$ $\sim$ 150 - 250 km s$^{-1}$) stays more constant with radius. Qualitatively, this trend of higher $\sigma_{*}$ galaxies displaying the largest radial IMF trends is in agreement with observations such as \cite{imf-martin-2}. Though, it is difficult to directly compare our results to observations by radius due to the differences in how effective radius ($R_{e}$) is measured and how the stellar half-mass radius is calculated for Illustris galaxies.

The decrement in IMF mismatch parameter, $\Delta$log($\alpha_{\rm IMF}$), and the maximum log($\alpha_{\rm IMF}$) of our highest velocity dispersion Illustris galaxies is similar to what is reported in \cite{imf-martin-2}. The decrement in log($\alpha_{\rm IMF}$) of the most bottom-heavy galaxy shown in Figure \ref{fig:radial_imf}, with $\sigma_{*}$ = 302 km s$^{-1}$, is $\Delta$log($\alpha_{\rm IMF}$) = 0.36 in cylinders from a galactocentric radius of (1/7)$R^{p}_{1/2}$ to 4$R^{p}_{1/2}$. Considering hollow cylinders in which log($\alpha_{\rm IMF}$) is calculated, from (1/7)$R^{p}_{1/2}$ to 4$R^{p}_{1/2}$ there is a larger decrement of $\Delta$log($\alpha_{\rm IMF}$) = 0.57. \cite{imf-martin-2} reports for their high velocity dispersion galaxy ($\sigma$ $\sim$ 300 km s$^{-1}$) a decrement of $\Delta$x = 1.15, which roughly corresponds to $\Delta$log($\alpha_{\rm IMF}$) = 0.35, from the center of the galaxy (r = 0 $R_{e}$) to 0.7 $R_{e}$.

\subsection{Scatter}
\label{sec:scatter}

As mentioned, increasing the slope of the input $\alpha_{\rm IMF}$ relation increases the scatter of the resulting overall log($\alpha_{\rm IMF}$)-$\sigma_{*}$ relations (Figure \ref{fig:global_imf} and Figure \ref{fig:imf}). This is consistent with observational studies that also show substantial scatter in the reported IMF-$\sigma$ relations \citep{imf-posacki,imf-conroy-1,imf-cappellari-1}. For example, based on dynamical modeling of ETGs in the ATLAS$^{\rm 3D}$ project, \cite{imf-cappellari-2} reports a 1$\sigma$ scatter of $\approx$0.12 dex (or 32\%) in their derived relation between IMF mismatch parameter and velocity dispersion.

\cite{imf-cappellari-2}'s reported scatter is comparable to the scatter seen in our Illustris log($\alpha_{\rm IMF}$)-$\sigma_{*}$ relations. Constructing the overall log($\alpha_{\rm IMF}$) of each galaxy based on the stellar velocity dispersion of each stellar particle's progenitor galaxy (Section \ref{sec:globalstar}), we produce a 1$\sigma$ scatter of 0.079 dex (20\%) using the original \cite{imf-spiniello} relation as input and a 1$\sigma$ scatter of 0.123 dex (32.7\%) using the 3.6$\times$ steeper input relation. Similarly, constructing the overall log($\alpha_{\rm IMF}$) using the local velocity dispersion of each stellar particle at the time of birth (Section \ref{sec:dmsigma}), we produce a 1$\sigma$ scatter of 0.045 dex (11\%) using the shallow input relation and a 1$\sigma$ scatter of 0.125 dex (33\%) using the steeper input relation that is able to reproduce the overall log($\alpha_{\rm IMF}$)-$\sigma_{*}$ relation. As discussed in Section \ref{sec:dmsigma}, the scatter  in our Illustris IMF relations is due to galaxies of similar global $z=0$ velocity dispersions having varying stellar birth property distributions like $\sigma_{\rm birth}$ or $\sigma_{*}$. These differences in the physical conditions of star-formation reflect an intrinsic scatter in the formation histories of galaxies with the same global $z=0$ properties.

\subsection{Redshift evolution}
\label{sec:redshift}

Finally, we examine the redshift evolution of the log($\alpha_{\rm IMF}$)-$\sigma_{*}$ relation by repeating the analysis described in Section \ref{sec:dmsigma} but now with a sample of galaxies at a higher redshift. In Illustris-1 at $z = 2$, we select the 311 galaxies with stellar masses greater than 10$^{10}$ M$_{\odot}$ and stellar velocity dispersions greater than 150 km s$^{-1}$. We do not place a cut on star-formation as we did for the $z = 0$ sample since fewer galaxies meet the sSFR \textless ~10$^{-11}$ yr$^{-1}$ criterion at $z = 2$. The average stellar mass, stellar velocity dispersion, and specific star formation rate of the $z = 2$ sample is $\overline{M_{*}}$ = 10$^{10.89}$, $\overline{\sigma_{*}}$ = 197 km s$^{-1}$, and $\overline{\rm sSFR}$ = 7.60\e{-10} yr$^{-1}$.

\begin{figure}
  \includegraphics[width=1.\columnwidth]{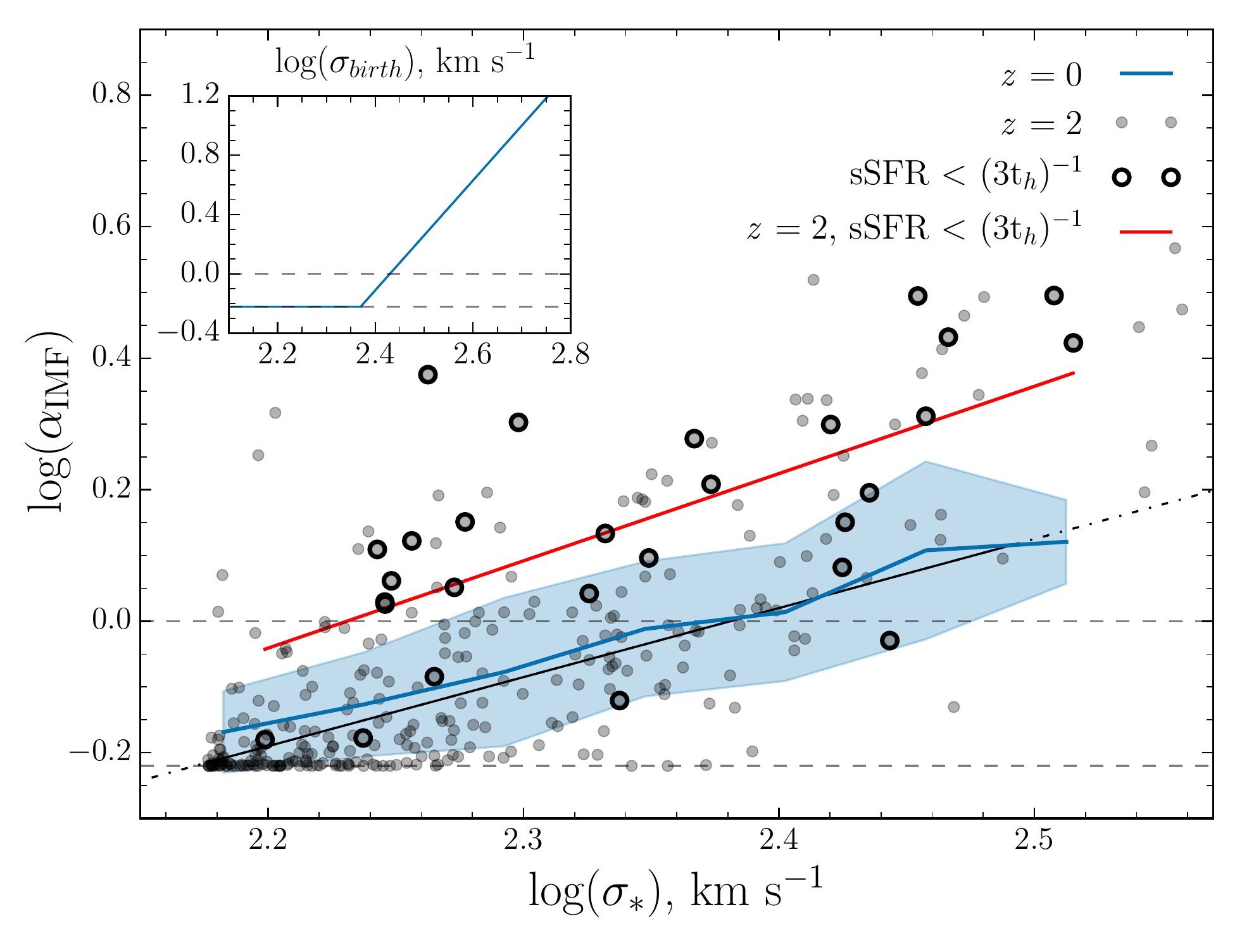}
    \caption{IMF mismatch parameter versus global velocity dispersion for the $z = 2$ massive galaxies. The grey points show the $z=2$ galaxies, while the blue relation shows the $z = 0$ log($\alpha_{\rm IMF}$)-$\sigma_{*}$ relation constructed using the steep input relation (shown in the inset as in Figure 3). The red line shows the fit to the $z = 2$ galaxies defined as quiescent with sSFR \textless ~6.82\e{-11} yr$^{-1}$ (outlined in black). At fixed velocity dispersion, quiescent galaxies at $z=2$ are \textit{more} bottom-heavy than their $z=0$ counterparts.
\medskip }
    \label{fig:redshift2}
\end{figure}

We construct the overall log($\alpha_{\rm IMF}$) of each $z = 2$ galaxy based on the local, birth velocity dispersion of the stellar particles following the same procedure outlined in Section \ref{sec:dmsigma}. Figure \ref{fig:redshift2} shows the resulting $z = 2$ log($\alpha_{\rm IMF}$)-$\sigma_{*}$ relation using the same steep input relation that was able to reproduce the observed log($\alpha_{\rm IMF}$)-$\sigma_{*}$ relation for the $z = 0$ galaxy sample. As seen in the figure, the overall log($\alpha_{\rm IMF}$) values of the $z = 2$ galaxies are, on average, higher than the $z = 0$ relation. Furthermore, the more quiescent galaxies generally have higher log($\alpha_{\rm IMF}$) values than the more star-forming galaxies. To more directly compare to the $z = 0$ relation, which only includes quiescent galaxies, we fit the $z = 2$ log($\alpha_{\rm IMF}$)-$\sigma_{*}$ relation only for galaxies with sSFR \textless ~(3t$_{h}$)$^{-1}$ yr$^{-1}$ where t$_{h}$ is the age of the Universe at a given redshift \citep{imf-damen}. In Figure \ref{fig:redshift2} the red line shows the fit to the 29 quiescent $z = 2$ galaxies, which is $\sim$1.4$\times$ steeper and offset by $\sim$0.17 dex compared to the $z = 0$ relation. 

The offset of the $z = 2$ log($\alpha_{\rm IMF}$)-$\sigma_{*}$ relation towards higher log($\alpha_{\rm IMF}$) values compared to the $z = 0$ relation is due to the assembly history of massive galaxies. In $\Lambda$CDM, massive galaxies are thought to first build up their in-situ stellar populations and then at later redshifts accrete smaller systems and build up their ex-situ stellar populations \citep{r-naab,r-oser}. In Illustris, quiescent galaxies at $z = 2$ have already formed a significant portion of their in-situ stellar particles, but have yet to accumulate a majority of their ex-situ stellar particles. The higher log($\alpha_{\rm IMF}$) values of the $z = 2$ galaxies suggests the stellar particles already belonging to galaxies by $z = 2$ are formed in higher velocity dispersion environments than the stellar particles that will be added to the galaxies at later times. To go from the $z = 2$ to the $z = 0$ log($\alpha_{\rm IMF}$)-$\sigma_{*}$ relation, stellar particles added to galaxies after $z = 2$ decrease the overall log($\alpha_{\rm IMF}$) values of the galaxies.

As will be discussed in Section \ref{sec:discussion}, our $z=2$ log($\alpha_{\rm IMF}$)-$\sigma_{*}$ relation is seemingly in tension with IMF observations beyond $z=0$, which suggest that the relation has remained constant over the past $\sim$8 Gyrs. Robust IMF determinations out to $z=2$ will be needed to fully assess the implications of our high redshift results.

%%%%%%%%%%%%%%%%%%%% RESOLUTION %%%%%%%%%%%%%%%%%%

\section{Dependence on resolution \& physics variations}
\label{sec:res_var}

\subsection{Convergence with resolution}
\label{sec:resolution}

First we confirm the convergence of our results to degradation in simulation resolution. We repeat the same analysis as described in Section \ref{sec:dmsigma} for Illustris-1 on the two lower resolution simulations, Illustris-2 and Illustris-3. All three simulations have the same box size of (106.5 Mpc)$^{3}$, but Illustris-1 contains $\sim$2 $\times$ 1820$^{3}$ resolution elements while Illustris-2 and Illustris-3 contain $\sim$2 $\times$ 910$^{3}$ and $\sim$2 $\times$ 455$^{3}$ resolution elements respectively. In Illustris-2 the average baryonic particle mass is $\overline{m_{b}}$ = 1.0\e{7} M$_{\odot}$ and in Illustris-3 it is $\overline{m_{b}}$ = 8.05\e{7} M$_{\odot}$. Refer to Table \ref{tab:illustris} for more Illustris-2 and Illustris-3 simulation parameters.

Implementing the same galaxy selection outlined in Section \ref{sec:selection} results in 229 galaxies in Illustris-2 and 103 galaxies in Illustris-3 that meet the three criteria at $z = 0$ of stellar mass, specific star formation rate, and velocity dispersion. The average stellar mass, specific star formation rate, and stellar velocity dispersion of the Illustris-2 sample is $\overline{M_{*}}$ = 10$^{11.44}$, $\overline{\rm sSFR}$ = 2.52\e{-12} yr$^{-1}$, and $\overline{\sigma_{*}}$ = 199 km s$^{-1}$ while for the Illustris-3 sample they are $\overline{M_{*}}$ = 10$^{11.48}$, $\overline{\rm sSFR}$ = 2.64\e{-12} yr$^{-1}$, and $\overline{\sigma_{*}}$ = 204 km s$^{-1}$. To determine the log($\alpha_{\rm IMF}$)-$\sigma_{*}$ relation we use both the observed relation as input (Equation \ref{eq:spiniello}) and the steeper log($\alpha_{\rm IMF}$)-$\sigma_{\rm birth}$ relation we constructed which was found to reproduce the observed trend with global velocity dispersion.

\begin{figure*}
  \includegraphics[width=2.1\columnwidth]{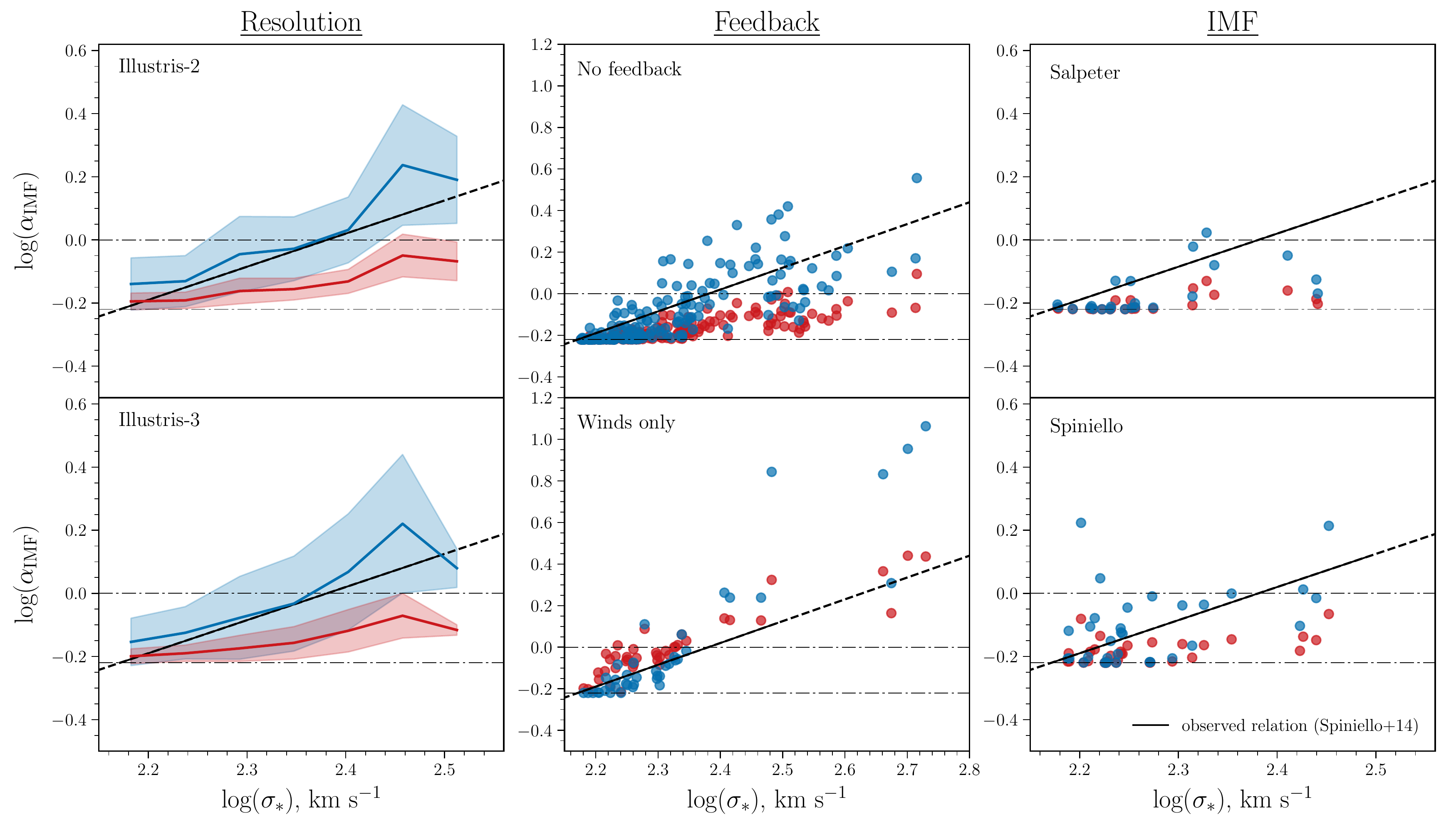}
      \caption{The log($\alpha_{\rm IMF}$)-$\sigma_{*}$ Illustris relations, constructed based on the birth velocity dispersion distributions of each galaxy, for the two lower resolution Illustris simulations (Illustris-2 and Illustris-3), the two simulations with varying feedback (no feedback and winds only), and the two simulations with varying IMFs (IMF-Salpeter and IMF-Spiniello). In each panel, the red curve or points show the observed relation (Equation \ref{eq:shifted}) used as input to construct the overall log($\alpha_{\rm IMF}$) of each galaxy and the blue curve shows the resulting relation using the steeper log($\alpha_{\rm IMF}$)-$\sigma_{\rm birth}$ relation. In each panel, the log($\alpha_{\rm IMF}$)-$\sigma_{*}$ constructed using just the stellar particles within 0.5$R^{p}_{1/2}$ is shown.
\bigskip }
    \label{fig:res_var}
\end{figure*}

The leftmost panels of Figure \ref{fig:res_var} show the resulting log($\alpha_{\rm IMF}$)-$\sigma_{*}$ trends for Illustris-2 and Illustris-3, constructed using just the birth velocity dispersions of the stellar particles within 0.5$R^{p}_{1/2}$ from the center of each galaxy. As with Illustris-1, using the observed relation to set the IMF of stellar particles at their birth times results in a $z = 0$ log($\alpha_{\rm IMF}$)-$\sigma_*$ relation that is shallower than observed (red curve). The Illustris-2 relation is $\sim$2.5$\times$ shallower than the observed relation while the Illustris-3 relation is $\sim$2.8$\times$ shallower than the observed relation. We construct the log($\alpha_{\rm IMF}$)-$\sigma_{*}$ using the same, steeper input relation we found to reproduce the global trend with Illustris-1, log($\alpha_{\rm IMF}$) = 3.7log($\sigma_{\rm birth}$) - 8.99. The blue curve in the figure shows that the same steep input relation that was able to match the global $\sigma_{*}$ trend in Illustris-1 is also able to reproduce the observed trend, within the uncertainties, in Illustris-2 and Illustris-3. Thus, we conclude our main results to be robust to resolution degradation.

\begin{figure}
  \includegraphics[width=1.\columnwidth]{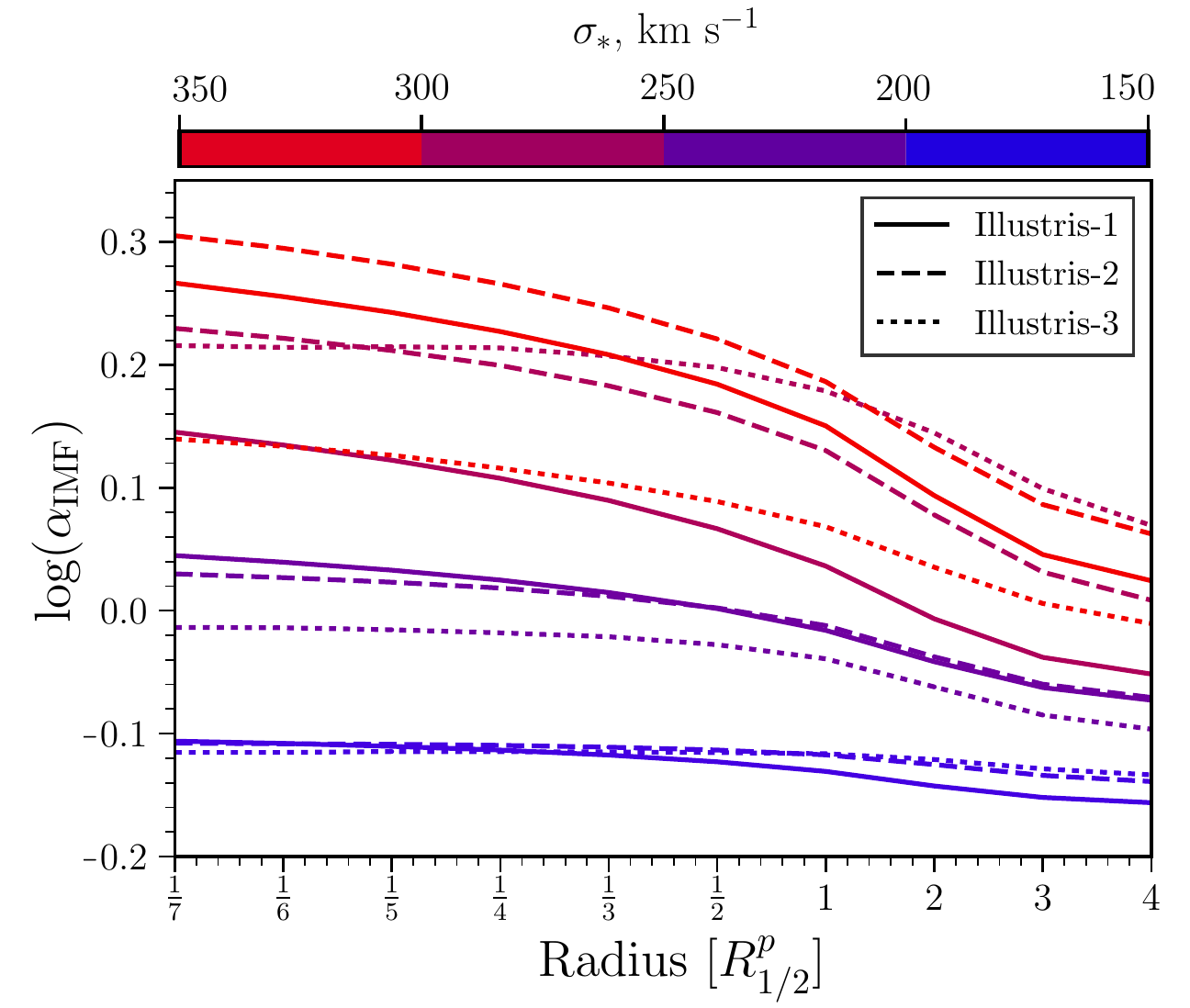}
    \caption{The average radial IMF profiles in four velocity dispersion bins for Illustris-1, -2, and -3. Each curve represents the average IMF mismatch parameter as a function of galactocentric radius, colored by velocity dispersion bin. For higher velocity dispersions, Illustris-3 exhibits shallower radial profiles at R$^{p}_{1/2}$ \textless ~1 compared to Illustris-1 and -2.
\medskip }
    \label{fig:radial_res}
\end{figure}

We also consider the effect of resolution on the radial IMF trends we explored in \ref{sec:radial}. Figure \ref{fig:radial_res} shows the average IMF profile (measured in cylinders) in four velocity dispersion bins for Illustris-1, Illustris-2, and Illustris-3. For the two lowest velocity dispersion bins, $\sigma_{*}$=150-200 km s$^{-1}$ and $\sigma_{*}$=200-250 km s$^{-1}$, the radial profiles for the three resolution levels are similar, although in the $\sigma_{*}$=200-250 km s$^{-1}$ bin the Illustris-1 profile is slightly steeper at R$^{p}_{1/2}$ \textless ~1 compared to Illustris-2 and -3. For the $\sigma_{*}$=250-300 km s$^{-1}$ bin, the Illustris-2 and -3 average profiles are at higher log($\alpha_{\rm IMF}$) values compared to Illustris-1. This is also seen comparing Figure \ref{fig:res_var} to Panel A  of Figure \ref{fig:imf}. In this velocity dispersion bin, the Illustris-3 radial profile is significantly shallower than the Illustris-1 and -2 profiles, which is due to the larger smoothing length of the Illustris-3 simulation. For the Illustris-2 and -3 simulations, the velocity dispersion bin with the highest log($\alpha_{\rm IMF}$) values is not the $\sigma_{*}$=300-350 km s$^{-1}$ bin, but the $\sigma_{*}$=250-300 km s$^{-1}$ bin. This is likely due to the most massive galaxies in the lower resolution simulations being more affected by intracluster light, which acts to reduce the overall log($\alpha_{\rm IMF}$). While the most relevant radial profile comparison would be between Illustris-1 and higher resolution simulations, as we discuss next, these high resolution simulations at the appropriate mass scale are not currently available.

Additionally, we attempt to test the robustness of our results to higher resolutions. This is motivated by \cite{r-sparre}, who present zoom-in simulations of Illustris galaxies with mass resolution up to 40 times better than that of Illustris-1. While galaxies in Illustris-1 do undergo nuclear starbursts \citep{ib-wellons-1}, in some of these zoom-in simulations the merger-driven nuclear starbursts are stronger. This could potentially influence the constructed IMF for two reasons. First, these starburst episodes produce more stellar mass. Second, these extreme star-formation environments display larger velocity dispersions. Hence they have the potential to be the sites where the bottom-heavy IMF of ETGs is built up.

These galaxies were selected by \citet{r-sparre} based on their $z=0$ quiescence and that they undergo a major merger between $z=.5$ and $z=1$. Each is run at three resolution levels: 1 - $m_{\rm dm}$ = 4.42\e{6} $M_{\odot}$, 2 - $m_{\rm dm}$ = 5.53\e{5} $M_{\odot}$, 3 - $m_{\rm dm}$ = 1.64\e{5} $M_{\odot}$. We focus on the two galaxies that exhibit the largest increase in star-formation rate at their respective merger times (galaxies 1349 and 1605). For the IMF analysis, we construct the overall log($\alpha_{\rm IMF}$) based on the $\sigma_{\rm birth}$ distributions of the stellar particles residing within 0.5$R^{p}_{1/2}$ of each respective galaxy. We find that the overall velocity dispersion of both galaxies increases with resolution level: $\sigma_{*}$ = 115, 123, and 134 km s$^{-1}$ for galaxy 1349 and $\sigma_{*}$ = 113, 123, and 139 km s$^{-1}$ for galaxy 1605. However, using the steep input relation as discussed in Section \ref{sec:dmsigma}, log($\alpha_{\rm IMF}$) hardly changes: -0.22, -0.22, -0.217 for galaxy 1349 and -0.22, -0.22, -0.22 for galaxy 1605, for zoom levels 1, 2, and 3, respectively. The reason is that the corresponding $\sigma_{\rm birth}$ distributions lie mostly below the Chabrier minimum of log($\alpha_{\rm IMF}$) = -0.22 set at $\sigma_{\rm birth}$ = 235 km s$^{-1}$. Using a shifted input relation so that fewer stellar particles are assigned the minimum Chabrier value, we do see more significant increases in log($\alpha_{\rm IMF}$) with increasing resolution level. This suggests the possibility that for zooms of higher velocity dispersion galaxies of at least $\sigma_{*}$ = 300 km s$^{-1}$ a shallower input relation might suffice to reproduce the observed log($\alpha_{\rm IMF}$)-$\sigma_{*}$ relation. However, such simulations would require a very significant investment of computing time and are currently not available. Hence, our zoom analysis is at this time inconclusive.

Lastly, we test the dependence of our results to time resolution by diluting the snapshots by a factor of 3 in our Illustris-2 analysis. We find our main result to be unaffected by this decrease in time resolution, justifying our choice for the number of snapshots produced in modified physics runs which will be discussed in the following section.

%%%%%%%%%%%%%%%%%%%% VARIATIONS %%%%%%%%%%%%%%%%%%

\subsection{Variations in simulation physics}
\label{sec:variations}

We also test the robustness of our results to variations in simulation physics, first considering variations in feedback. We ran a box of 40 Mpc/h on a side with $2\times320^{3}$ resolution elements (with a Chabrier IMF) once with no feedback and once with galactic winds but no AGN feedback. In each simulation we select galaxies with stellar masses greater than 10$^{10}$ M$_{\odot}$ and stellar velocity dispersions greater than 150 km s$^{-1}$. Since no galaxies in the winds only simulation meet the sSFR criterion, we do not cut on sSFR. This selection results in 178 $z = 0$ galaxies in the no feedback simulation and 41 $z = 0$ galaxies in the winds only simulation. The average stellar mass, specific star formation rate, and stellar velocity dispersion of the no feedback sample is $\overline{M_{*}}$ = 10$^{11.51}$, $\overline{\rm sSFR}$ = 2.68\e{-11} yr$^{-1}$, and $\overline{\sigma_{*}}$ = 216 km s$^{-1}$, while for the winds only sample $\overline{M_{*}}$ = 10$^{11.43}$, $\overline{\rm sSFR}$ = 1.74\e{-10} yr$^{-1}$, and $\overline{\sigma_{*}}$ = 223 km s$^{-1}$.

The middle panels of Figure \ref{fig:res_var} show the resulting log($\alpha_{\rm IMF}$)-$\sigma_{*}$ relations constructed within 0.5$R^{p}_{1/2}$ for the no feedback simulation (top) and winds only simulation (bottom), where the overall log($\alpha_{\rm IMF}$) of each galaxy is calculated based on the local, birth velocity dispersions of the stellar particles. For the no feedback simulation, the shallow input relation (as used in Section \ref{sec:dmsigma}) produces a log($\alpha_{\rm IMF}$)-$\sigma_{*}$ relation that is shallower than the observed relation, similar to the corresponding Illustris-1 relation. We find that the same steep input relation that was necessary to reproduce the observed log($\alpha_{\rm IMF}$)-$\sigma_{*}$ trend for Illustris-1 is also able to reproduce the observed trend in the no feedback simulation. The winds only simulation, on the other hand, varies from the Illustris and no feedback results. As seen in bottom, middle panel of Figure \ref{fig:res_var}, the shallow input relation \textit{does} reproduce the observed log($\alpha_{\rm IMF}$)-$\sigma_{*}$ relation while the steeper input relation results in a log($\alpha_{\rm IMF}$)-$\sigma_{*}$ relation that is $\sim$2$\times$ too steep. 

The impact of varying the simulation feedback on the overall log($\alpha_{\rm IMF}$) calculated for each galaxy can be understood by considering what star-formation is suppressed. AGN feedback suppresses star-formation in massive galaxies through both quasar and radio mode, while galactic winds suppress star-formation in lower mass galaxies with shallower potentials. In the no feedback simulation, without AGN feedback or galactic winds, star-formation is not suppressed either in low- or high-mass galaxies. The fraction of high to low velocity dispersion stellar particles in the no feedback simulation ends up being similar to the fraction in the full physics Illustris simulations, and a steep input relation is needed to reproduce the observed log($\alpha_{\rm IMF}$)-$\sigma_{*}$ trend. In the winds only simulation, star-formation in low-mass galaxies is suppressed but not in high-mass galaxies. The fraction of high to low velocity dispersion stellar particles in enhanced, leading to galaxies having higher log($\alpha_{\rm IMF}$) values. Since more massive galaxies have an increasing fraction of stellar particles formed ex-situ in lower-mass galaxies, an increasing fraction of low velocity dispersion stellar particles are suppressed, allowing the shallow input relation to reproduce the slope of the log($\alpha_{\rm IMF}$)-$\sigma_{*}$ trend. Of the physical quantities and simulation variations explored in this paper, a stellar mass function tilted in favor of star-formation in high-mass galaxies is able preserve the overall IMF trend without requiring a steep, physical IMF relation. Although, indicated by the lack of massive quiescent galaxies, the galaxy population in the winds only simulation is unrealistic. As shown in \cite{ib-vogelsberger-3}, not including AGN feedback results in a $z=0$ stellar mass function and stellar mass to halo mass relation that are too high compared to observations and the fiducial Illustris model, as well as a star-formation rate density and a stellar mass density functions with redshift also being too high. 

Lastly, we begin to explore the effect of varying the IMF with which the simulation is run. As discussed in Section \ref{sec:illustris}, a Chabrier IMF law is used to govern mass and metal return from stellar particles in Illustris, as well as to calculate the mass-loading factors of galactic winds. Our empirical approach to constructing the IMF mismatch parameter of Illustris galaxies is not expected to depend directly on the IMF used to run the simulation. But, different mass returns and feedback may affect the local velocity dispersions which stellar particles are born into. To test if the IMF that the simulation is run with alters our main result, we run simulations with different IMF laws but otherwise with the same physics models as in Illustris, using smaller boxes at lower resolutions. We run a box of 40 Mpc/h on a side with $2\times320^{3}$ resolution elements incorporating two IMF laws: 1) a pure Salpeter IMF law with a slope of $x = 2.35$ and 2) the variable IMF law presented by \cite{imf-spiniello} dependent on the local dark matter velocity dispersion $\sigma_{\rm birth}$. More details about these additional simulations are listed in Table \ref{tab:illustris}. 

Using the same galaxy selection criteria outlined in Section \ref{sec:selection}, 23 galaxies in the Salpeter simulation at $z = 0$ and 31 galaxies in the Spiniello simulation at $z = 0$ meet the stellar mass, specific star formation rate, and velocity dispersion criteria. The average stellar mass, specific star formation rate, and stellar velocity dispersion of the Salpeter sample is $\overline{M_{*}}$ = 10$^{11.34}$, $\overline{\rm sSFR}$ = 1.27\e{-12} yr$^{-1}$, and $\overline{\sigma_{*}}$ = 190 km s$^{-1}$, while for the Spiniello sample $\overline{M_{*}}$ = 10$^{11.38}$, $\overline{\rm sSFR}$ = 3.94\e{-13} yr$^{-1}$, and $\overline{\sigma_{*}}$ = 189 km s$^{-1}$. Again, we determine the log($\alpha_{\rm IMF}$)-$\sigma_{*}$ relation using both the observed relation as input (Equation \ref{eq:shifted}) and the steeper log($\alpha_{\rm IMF}$)-$\sigma_{\rm birth}$ relation we constructed which was found to reproduce the observed trend with global $\sigma_{*}$ in the Illustris-1 analysis.

The rightmost panels of Figure \ref{fig:res_var} show the resulting log($\alpha_{\rm IMF}$)-$\sigma_{*}$ relations constructed within 0.5$R^{p}_{1/2}$ for our Salpeter and Spiniello IMF runs. Consistent with the results in Section \ref{sec:dmsigma}, for both modified IMF simulations the output log($\alpha_{\rm IMF}$) values constructed using the steeper log($\alpha_{\rm IMF}$)-$\sigma_{\rm birth}$ input relation lie closer to the observed relation than the log($\alpha_{\rm IMF}$) values constructed with the shallow input relation. While there are a few outliers and the number of high velocity dispersion galaxies in each simulation is small, incorporating a different and even variable IMF in the Illustris galaxy formation model does not seem to significantly modify the results of this paper. Since the formal fits of the Salpeter and Spiniello output log($\alpha_{\rm IMF}$)-$\sigma_{*}$ relations are shallower than the observed relation, this serves to strengthen our claim that a steeper input relation is required to produce the observations of IMF variations. Lastly, while the galaxy population in these modified IMF simulations undoubtedly differs from the galaxy population in Illustris, the investigation of these differences and the implications of simulations which self-consistently include a variable IMF is the topic of future work.

%%%%%%%%%%%%%%%%%%% DISCUSSION %%%%%%%%%%%%%%%%%%%

\section{Discussion}
\label{sec:discussion}

\subsection{Simulation limitations}

Before we discuss the possible implications of this work, it is important to reiterate the extent to which the results are dependent on the simulation models and resolution. While the Illustris galaxy formation and evolution models reproduce many key galaxy properties and scaling relations, there is certainly room for improvement \citep{ib-vogelsberger-a, ib-vogelsberger-b, ib-genel}. For example, Illustris produced massive galaxies with too high of a stellar mass \citep{ib-vogelsberger-b}, and galaxy sizes that are too large \citep{pillepich}. In particular, relevant to studying IMF variations through the hierarchical build-up of galaxies, is the merger rate. \cite{ib-merger} show the Illustris merger rate to match some observations well, though there remain qualitative differences among the observations. If the merger rate in Illustris is too high, this might explain why a steep physical IMF relation is needed to conserve the global IMF relation to $z=0$.

Another simulation parameter that could be influencing our results is resolution. Observational IMF studies are starting to reveal that IMF variations are confined to the most inner parts of galaxies, at radii typically \textless ~0.3R$_{e}$ \citep{imf-martin-2, imf-vd}. The resolution of the Illustris simulation, with the highest resolution simulation having a baryonic smoothing length of 0.7 kpc, prevents us from gaining a realistic understanding of IMF gradients at small radii. As considered in Section \ref{sec:resolution}, higher resolution simulations could yield more powerful nuclear starbursts, which in turn could reduce the steepness of the physical IMF laws we currently find necessary. At this time though, there are no simulations at the relevant mass scale to test this hypothesis.

Given the above dependence on simulation details, it is advised that the results of this work be interpreted in a more qualitative sense. The exact input relation we find necessary to reproduce the observed IMF trends are certainly sensitive to the galaxy formation and evolution model used, as well as resolution. While the quantitative results may change with new and improved simulation, the qualitative results are more robust.

\subsection{Implications for IMF observations}
\label{sec:obs}

Observations suggest that with increasing galaxy velocity dispersion, galaxies also have increasingly bottom-heavy IMFs where the most massive $z = 0$ galaxies are characterized by super-Salpeter IMF slopes. Since these massive galaxies are believed to be primarily composed of ex-situ stellar populations, which formed in smaller systems with lower velocity dispersions and only later accreted onto the main galaxy, then the physical explanation for the steepness of the correlation between $z = 0$ global galaxy properties and the IMF is unclear. The result that even steeper physical IMF relations are needed to preserve the observed IMF variations through the assembly of massive galaxies has implications for observations of IMF variations and theoretical work predicting IMF variations.

One consequence of our analysis is that massive, quiescent galaxies at higher redshifts would have global IMFs even more bottom heavy than their $z = 0$ counterparts. This is because massive galaxies in Illustris first build up their in-situ stellar populations, which are mainly formed in high velocity dispersion environments, and only later accrete smaller systems with stellar populations formed in low velocity dispersion environments. For massive galaxies, these smaller systems reside even within an effective radius where global IMF measurements are typically made \citep{r-gomez}, reducing the overall IMF of the galaxy. 

Based on studies of ETG populations at $z \sim 1$ and $z \sim 1.4$, observations beyond the local Universe are beginning to suggest that the IMF-$\sigma$ relation remains roughly constant over the last 8 Gyrs \citep{imf-martin-4, imf-gargiulo-2}. Contrary to these studies, \cite{imf-sonnenfeld-0} find that the overall log($\alpha_{\rm IMF}$) at fixed velocity dispersion decreases from the $z \sim 0$ value out to $z \sim 0.8$. In \cite{imf-sonnenfeld}, they postulate that a possible source of this apparent evolution could be their assumption of fixed dark matter density profile. A robust determination of the evolution of IMF-$\sigma$ relation from $z \sim 0$ to $z \sim 2$ will require careful determination of the dark matter fraction of galaxies to break the dark matter-IMF degeneracy. To reconcile our prediction for $z \sim 2$ with observations that suggest a constant IMF-$\sigma$ evolution, one option is to invoke a time-dependent physical IMF `law'. In the context of our analysis, this would mean applying a shallower log($\alpha_{\rm IMF}$)-$\sigma_{\rm birth}$ relation to higher-redshift stellar population.

\begin{figure}
  \includegraphics[width=1.0\columnwidth]{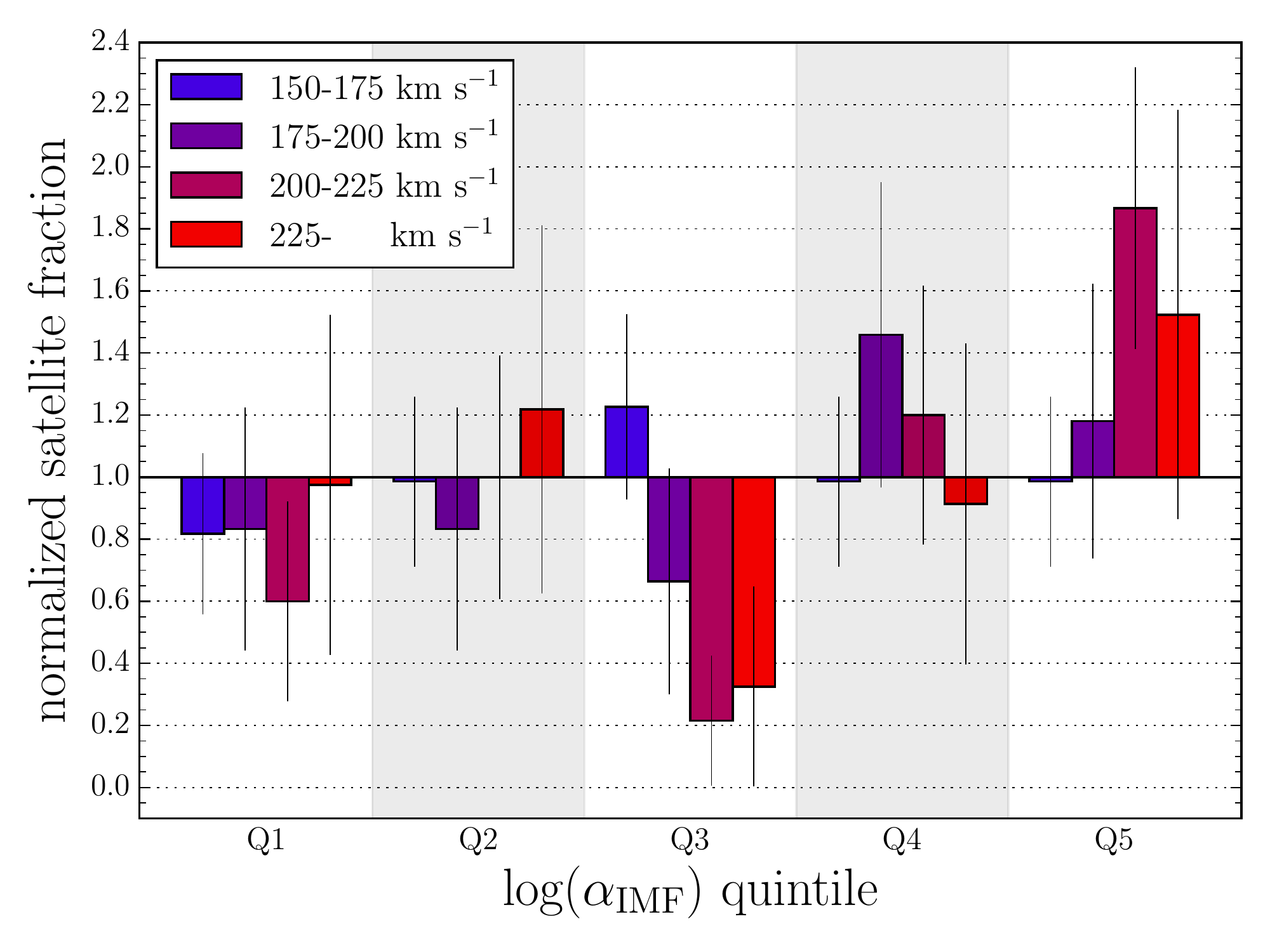}
      \caption{The satellite fraction of $z=0$ Illustris-1 galaxies in 4 velocity dispersion bins, normalized by the total satellite fraction in the bin, for each log($\alpha_{\rm IMF}$) quintile. The bars are colored according to velocity dispersion, and each group of bars shows one quintile. The errors are calculated assuming a binomial distribution, where we add in quadrature the error associated with the total satellite fraction and the error associated with the satellite fraction in each quintile. In the 200-225 km s$^{-1}$ bin, satellite galaxies make up 65\% of the most extreme log($\alpha_{\rm IMF}$) galaxies, which is $\sim$1.8$\times$ larger than their total fraction in the bin.
\medskip }
    \label{fig:satellite}
\end{figure}

A further prediction of our analysis is that satellite galaxies, at fixed velocity dispersion, should have more bottom-heavy IMFs than central galaxies. Since satellites are expected to undergo fewer minor mergers than centrals, they should better preserve to $z=0$ their bottom-heavy IMFs that are in place at higher redshift before they become satellites. To demonstrate this prediction, we examine the satellite fraction for $z=0$ Illustris-1 galaxies in four velocity dispersion bins, with widths of 25 km s$^{-1}$ except for the highest $\sigma_{*}$ bin that includes all 78 galaxies with $\sigma_{*}$ \textgreater ~225 km s$^{-1}$. For each velocity dispersion bin, we determine log($\alpha_{\rm IMF}$) quintiles and calculate the fraction of satellite galaxies within each quintile. Figure \ref{fig:satellite} shows the satellite fraction, normalized by the total satellite fraction in the velocity dispersion bins, for each log($\alpha_{\rm IMF}$) quintile. We note that the formal errors associated with the normalized satellite fractions are large, especially when comparing different velocity dispersion bins within each quintile. However, some individual bins are above or below the total satellite fraction with statistical significance.

The fifth quintile represents the galaxies with the most bottom-heavy IMFs in each velocity dispersion bin. For the lowest velocity dispersion bin, from 150 - 175 km s$^{-1}$, the satellite fraction is the same as the total satellite fraction in the bin, indicating that for these low velocity dispersion galaxies, satellites do not have higher log($\alpha_{\rm IMF}$) values than centrals. This can be understood as lower velocity dispersion galaxies are more dominated by in-situ evolution and less affected by minor mergers. In higher velocity dispersion bins, the fraction of satellites composing the most extreme log($\alpha_{\rm IMF}$) galaxies is higher than the total satellite fraction. In particular, in the 200-225 km s$^{-1}$ bin $\sim$65\% of the highest log($\alpha_{\rm IMF}$) galaxies are satellites, which is significantly enhanced with respect to their total fraction in the bin. For these galaxies with 200-225 km s$^{-1}$, the highest IMF values of individual galaxies are log($\alpha_{\rm IMF}$) $\sim$ 0.32 (compared with the typical log($\alpha_{\rm IMF}$) $\sim$ -0.05 for this $\sigma_{*}$ bin), and a majority of these `extreme IMF' galaxies are satellites. For galaxies with $\sigma_{*}$ \textgreater ~225 km s$^{-1}$, of the 8 galaxies with log($\alpha_{\rm IMF}$)  \textgreater ~0.20, 2 are satellites and 6 are centrals. The most extreme log($\alpha_{\rm IMF}$) value for a central in this bin is log($\alpha_{\rm IMF}$) = 0.32 and for a satellite it is log($\alpha_{\rm IMF}$) = 0.36.

\subsection{Implications for IMF theory}

In our post-processing analysis of Illustris, we find that steep physical IMF relations, as applied to the birth properties of stellar particles, are required to reproduce the observed $z = 0$ IMF trend with global velocity dispersion. Input relations more than 3$\times$ steeper than the observed relation are needed, which means that some individual stellar populations must be formed with mass-to-light ratios up to $\sim$20$\times$ greater than the Salpeter mass-to-light ratio. These required extreme mass-to-light ratios are $\sim$10$\times$ greater than the overall mass-to-light ratios measured in observations of massive galaxies. 

To gain an idea of what IMF slope could give rise to an $M_{*}/L$ ratio excess this large, we calculate mass-to-light ratios with the \texttt{FSPS} (Flexible Stellar Population Synthesis) library \citep{fsps-1, fsps-2} and the Python \texttt{FSPS} package\footnote{http://dan.iel.fm/python-fsps/}. We model a single burst of star-formation with solar metallicity and an exponentially declining star-formation history truncated at 4 Gyrs, and calculate the $r$-band $M_{*}/L$ ratio at an age of 10 Gyr for several IMF slopes. We find that unimodal IMF slopes greater than $x=4$ are required to produce a $M_{*}/L$ ratio that is $\sim$20$\times$ greater than the Salpeter $M_{*}/L$ ratio, where an IMF slope of $x=4$ results in a $M_{*}/L$ excess of $\sim$11.5 and an IMF slope of $x=4.5$ results in an $M_{*}/L$ excess of $\sim$24.

While IMF slopes this steep have yet to be robustly observed, an IMF slope this extreme could have more of an immediate implication for analytical IMF variation theories. For example, the functional form of the CMF could be mapped to a unimodal IMF slope to determine how high of a Mach number would be required to produce these extreme $M_{*}/L$ ratios in theories that predict that high Mach number environments promote a bottom-heavy IMF \citep{r-hopkins, imf-chabrier-3, imf-gus}.

Additionally, in Section \ref{sec:sfr} we determined the overall log($\alpha_{\rm IMF}$)-$\sigma_{*}$ relation with an input relation based on star-formation rate. Motivated by IGIMF theory, we assigned stellar particles born into low SFR galaxies a higher log($\alpha_{\rm IMF}$) value and stellar particles born into high SFR lower log($\alpha_{\rm IMF}$) values. With this input relation, we are able to reproduce the slope and normalization of the observed log($\alpha_{\rm IMF}$)-$\sigma_{*}$ relation, but in the opposite direction. Combined with our ability to reproduce the slope of the observed relation and its sign with an input relation based on $\sigma_{\rm birth}$, this suggests that stellar particles formed in high SFR environments also form in high velocity dispersion environments. If this is the case, then there is an apparent tension between the analytical IMF theories which predict that high Mach number (and therefore high velocity dispersion) environments promote a bottom-heavy IMF \citep{r-hopkins, imf-chabrier-3, imf-gus}, and IGIMF theory \cite{imf-weidner-2} which predicts that high SFR environments lead to a top-heavy IMF. 

While this tension seems to exist, the main driver of IMF variations in analytical and simulation work is far from settled. For example, recent SPH simulations of star-formation \cite{imf-motta} show no correlation between Mach number and peak of the IMF in high density environments. Moreover, in low density environments they actually find that a higher Mach number shifts the peak of the IMF to higher masses, implying a more \textit{top-heavy} IMF -- opposite of what \cite{r-hopkins} predicts. In general, if observations prove to be robust, any theory predicting IMF variations will have to accommodate both the correlation of overall IMF with global $z=0$ velocity dispersion and with metallicity. One possibility is an IMF `law' which depends on a combination of local velocity dispersion and metallicity. Combined, this could be an additional probe of Mach number, with metallicity controlling cooling and therefore the sound speed.

As discussed in \cite{imf-weidner-2}, to resolve the tension between IGIMF theory and the observed mass excess of ETGs it is suggested that at least part of the inferred mass is due to an increased number of stellar remnants. IMF variation studies focused on measuring dwarf sensitive spectral lines do break the degeneracy between low-mass stars and stellar remnants \cite[e.g.][]{imf-vandokkum-1,imf-labarbera-1}, suggesting that the inferred, excess mass is low-mass stars. However, the exact parameterization of the IMF of massive ETGs, especially below 1 M$_{\odot}$, is still unconstrained. Determining the physical driver of IMF variations will require both theoretical work predicting the shape of the IMF as a function of environment and the ability to observationally constrain the shape of the IMF. Recent work to constrain the shape of low-mass end of the IMF from high-quality spectra appears promising \citep{imf-new-conroy}.

\subsection{Comparison to other work}

The closest approach in the existing literature to the post-processing IMF analysis of Illustris presented in this paper is that of \cite{imf-sonnenfeld}. For a sample of galaxies at $z=2$, \cite{imf-sonnenfeld} assume that all subsequent stellar mass growth occurs via dry mergers, based on a toy merging model. By using empirical log($\alpha_{\rm IMF}$)-$\sigma$ or log($\alpha_{\rm IMF}$)-$M_{*}$ relations, a log($\alpha_{\rm IMF}$) value is assigned to each central galaxy at $z = 2$. The overall log($\alpha_{\rm IMF}$) of each galaxy at later redshifts is determined by the addition of smaller systems, with their own IMFs assigned by the same empirical relations. 

While \cite{imf-sonnenfeld} explore the mixing between IMF dependence on stellar mass and velocity dispersion, the most directly comparable result to this work is the log($\alpha_{\rm IMF}$)-$\sigma$ relation based on the velocity dispersion model. Here they find that the slope and normalization of the log($\alpha_{\rm IMF}$)-$\sigma$ relation is preserved from $z=2$ to $z=0$, such that galaxies at fixed velocity dispersion have the same log($\alpha_{\rm IMF}$) at different times. As shown in Section \ref{sec:redshift}, our analysis predicts that at fixed velocity dispersion log($\alpha_{\rm IMF}$) is higher at $z=2$ than at $z=0$. These qualitatively discrepant results could be due to a number of factors stemming from significant methodology differences.

First, we assign a log($\alpha_{\rm IMF}$) value to stellar particles based on the \textit{birth} velocity dispersion, whereas in \cite{imf-sonnenfeld} log($\alpha_{\rm IMF}$) values are assigned based on the $z=2$ velocity dispersion for the central galaxies and based on later redshift values for the smaller systems. This difference could be why our analysis required steeper input relations to reproduce the observed relation in the first place, as birth velocity dispersion values are generally lower than at $z=2$ and later. Another difference between the works is that the \cite{imf-sonnenfeld} model does not take into account the addition of newly quenched galaxies to the population after $z=2$. This certainly affects the comparison to our $z=0$ IMF-$\sigma$ relation, which does include more recently quenched galaxies. Finally, as recognized by \cite{imf-sonnenfeld}, the pure dry merger model does not correctly reproduce the redshift evolution of the velocity dispersion of galaxies, suggesting that a dissipational component is missing. The addition of such a component to their model would also break the constancy of the log($\alpha_{\rm IMF}$)-$\sigma$ relation. On the other hand, our analysis is based on the innermost region of each galaxy, following observational constraints, while \cite{imf-sonnenfeld} do not address radial gradients. When we consider the full galaxy extent, we find an even stronger shallowing of the log($\alpha_{\rm IMF}$)-$\sigma$ relation with redshift, suggesting the differences between the two studies are even larger than at face value.

As mentioned in the Introduction, IMF variations have been studied in the context of semi-analytical models. In particular, \cite{imf-fontanot} incorporates IGIMF theory into the \texttt{GAEA} (GAlaxy Evolution and Assembly) model to study the implications of IMF variations on the chemical evolution and dynamical properties of galaxies. There, a broken power-law IMF is employed where for stellar masses less than $\sim$1$M_{\odot}$ the standard Kroupa IMF is adopted, and for masses greater than that the slope of the IMF is determined by the instantaneous star-formation rate. In accordance with IGIMF theory, higher star-formation rates correspond to shallower (more top-heavy) IMF slopes at the high-mass end. 

For $z=0$ galaxies formed under the IGIMF model, \cite{imf-fontanot} compares the stellar mass-to-light ratio ($M_{*}/L$) and stellar mass excesses to the corresponding Chabrier equivalents. Overall, their results are in qualitative agreement with observations of IMF variations, where higher mass galaxies exhibit a greater mass (and $M_{*}/L$) excess compared to the mass (or $M_{*}/L$) derived assuming a Chabrier IMF. But for 10$^{9}$ $M_{\odot}$ \textless ~$M_{*}$ \textless ~10$^{10.8}$ $M_{\odot}$, there is a negative slope to the `true' versus Chabrier stellar mass relation, which only turns over and becomes positive for stellar masses 10$^{10.8}$ $M_{\odot}$ \textless ~$M_{*}$ \textless ~ 10$^{12.2}$ $M_{\odot}$. In contrast, the overall IMF-$\sigma_{*}$ found in this study (Section \ref{sec:sfr}), assuming a SFR dependent IMF relation where higher SFRs correspond to a less bottom-heavy IMF, exhibits an overall negative slope from the lowest to highest velocity dispersion galaxies. While this tension could possibly be highlighting an interesting difference between the Illustris and \texttt{GAEA} galaxy formation models, there are several factors that make the comparison difficult including the post-processing nature of our analysis and our different parameterization of the IMF-SFR relation. 

Concerning the chemical enrichment of galaxies, \cite{imf-fontanot} find that the IMF variations they include are actually able to reproduce the observed $\alpha$-enhancement of massive galaxies ($M_{*}$ \textgreater ~10$^{11}$ $M_{\odot}$), though they do note that IMF variations are not the only solution to produce the observed metallicity of galaxies. Adopting a standard universal IMF, the [O/Fe] ratio of galaxies in \texttt{GAEA} with $M_{*}$ = 10$^{12}$ $M_{\odot}$ is $\sim$0.2 dex too low compared to observations of massive galaxies. At the time, the post-processing nature of our analysis prevents us from making a meaningful comparison to this result. In future work, as discussed in the next section, we will self-consistently incorporate different IMF laws into the Illustris galaxy formation models, and by tracking the build-up of individual metals we will be able to study how the chemical enrichment of galaxies is altered.

\subsection{Prospects for future work}
Future work will include validating our results with both higher resolution simulations, and improved galaxy formation models \citep{pillepich, rainer}. High-resolution simulations of individual galaxies, at the relevant mass scales, will allow for a more realistic study of radial gradients in the IMF and fully capture the impact of star-formation occurring in nuclear starbursts. Additionally, improved galaxy formation models that can better match observed galaxy properties than the current fiducial Illustris model can, such as the $z = 0$ stellar mass function, will be a crucial test of the robustness of our main result.

In addition to validating our results with improved simulations, future work will include expanding on our analysis. Inspired by observations and theoretical work, in this study we examined five physical quantities associated with star-formation and/or IMF variations. There are other quantities not considered here that have also been advocated to influence the IMF, including pressure and redshift \citep{imf-krumholz, imf-munshi}. Combinations of quantities can also be considered, such as metallicity and velocity dispersion. 

Expanding on the analysis of the variable IMF simulations presented in Section \ref{sec:variations}, further work is also required for studying the impact of a variable IMF that is incorporated self-consistently on the evolution and properties of galaxy populations. Since the number of low- to high-mass stars in a galaxy determines the amount of metals injected into the ISM, the energy available for supernovae feedback, and the mass of baryons trapped in low-mass stars, the inclusion of a variable IMF could significantly alter our current picture of galaxy evolution. While a variable IMF remains controversial because measuring the IMF is observationally difficult, the numerous studies reporting IMF variations and the analytical work predicting IMF variations in extreme star-formation environments necessitates that the implications of a variable IMF be investigated.

%%%%%%%%%%%%%%%%%%% CONCLUSION %%%%%%%%%%%%%%%%%%%

\section{Summary \& Conclusion}
\label{sec:conclusion}

In this study, we present an investigation of the physical origin of IMF variations using the cosmological simulation Illustris in post-processing. For a sample of massive ($M_{*}$ \textgreater ~10$^{10}$ $M_{\odot}$) and quiescent (sSFR \textless ~10$^{-11}$ yr$^{-1}$) galaxies we connect the physical conditions in which stellar particles form to the properties of the galaxies they reside in at $z = 0$. We do this by constructing the overall IMF mismatch parameter, $\alpha_{\rm IMF}$, of each galaxy based on various formation conditions associated with the individual stellar particles that comprise it. By attempting to reproduce the observed relations between overall IMF and global velocity dispersion, we are able to gain insight into how galaxy-wide quantities at $z = 0$ are related to the IMF -- a local property defined at the time of star-formation. Our findings are summarized as follows:

\begin{itemize}[leftmargin=7mm]

\item A  much steeper than observed physical IMF relation is needed to reproduce the reported IMF trends with global $z = 0$ velocity dispersion under the hierarchical assembly of massive galaxies. This result ties observations of an IMF that varies with $z = 0$ galactic properties to a physical origin, but requires some individual stellar populations to be formed with super-Salpeter IMFs that are even steeper than the IMFs reported by observational studies. These extreme IMFs are up to $\sim$20$\times$ in excess of the Salpeter $M_{*}/L$, which could imply a unimodal IMF slope of $x$ \textgreater ~4. 

\item Of the five physical quantities we consider, we are able to reproduce the observed IMF trend with $z = 0$ velocity dispersion by constructing the overall $\alpha_{\rm IMF}$ of each galaxy based on the global and local birth velocity dispersion of the stellar particles, and the global star-forming gas velocity dispersion of the progenitor galaxy in which each stellar particle was formed. All of these quantities are roughly related to Mach number, which in analytical models of IMF variations corresponds to increased fragmentation on lower mass scales in more extreme star-formation environments.  

\item We are unable to reproduce the observed IMF trend with $z=0$ velocity dispersion when constructing the overall $\alpha_{\rm IMF}$ of each galaxy based on the metallicities of individual stellar particles. The relations obtained in this way are too shallow. This is due to the scatter and near flatness of the [M/H]-$\sigma_{*}$ relation, and to the fact that the global metallicity of each galaxy is composed of a broad distribution of individual stellar particle metallicities. 

\item Using the star-formation rate of the progenitor galaxy in which each individual stellar particle was formed to construct the overall $\alpha_{\rm IMF}$, we obtain steep relations between IMF and $z=0$ velocity dispersion, but in the opposite direction to what is observed. Inspired by IGIMF theory, we construct a log($\alpha_{\rm IMF}$)-SFR relation in which a low SFR corresponds to a bottom-heavy IMF. For the $z=0$ massive quiescent galaxies we focus on in this study, this simple relation does not reproduce the direction of the observed IMF trend. If stellar particles which form in merger induced starbursts have both high velocity dispersions and high SFRs, there is tension between IMF theories as to whether these stellar populations are expected to form with a bottom-heavy or top-heavy IMF. 

\item We find radial gradients in the constructed log($\alpha_{\rm IMF}$) for massive galaxies due to the different formation conditions of stellar particles that reside at the center of galaxies versus in the outer regions. This result reinforces the need for consistent comparisons of IMF variations across observational studies that may be probing the IMF at different radii. It also further supports the idea that the IMF is a local property of galaxies, which are composites of numerous stellar populations formed in a diverse range of physical conditions throughout cosmic time. 

\item The scatter in the constructed Illustris IMF relations is reflective of the diverse formation histories of galaxies that have similar $z = 0$ velocity dispersions. In other words, galaxies with similar galactic quantities can be comprised of stellar populations which formed in different physical environments. This is in agreement with, and can provide a possible explanation for, the scatter in the observed IMF-$\sigma{*}$ relations. 

\item Based on our analysis we make two predictions for observations: (1) the log($\alpha_{\rm IMF}$)-$\sigma_{*}$ relation at high redshift would be more bottom-heavy than the $z=0$ relation, and (2) at high velocity dispersions galaxies with extreme bottom-heavy IMFs are preferentially satellite galaxies.

\end{itemize}

\section*{Acknowledgements}

We thank Richard Bower and Romain Teyssier for useful discussions, and Martin Sparre for sharing with us the outputs of his zoom-in simulations. We also thank the anonymous referee for a helpful report. KB is supported by the NSF Graduate Research Fellowship under grant number DGE 16-44869. SG acknowledges support provided by NASA through Hubble Fellowship grant HST-HF2-51341.001-A awarded by the STScI, which is operated by the Association of Universities for Research in Astronomy, Inc., for NASA, under contract NAS5-26555. GB acknowledges financial support from NASA grant NNX15AB20G and NSF grant AST-1615955. The Flatiron Institute is supported by the Simons Foundation. The Illustris simulation was run on the CURIE supercomputer at CEA/France as part of PRACE project RA0844, and the SuperMUC computer at the Leibniz Computing Centre, Germany, as part of project pr85je. Modified simulations used in this study were run on Columbia University's High Performance Computing cluster Yeti and on the Odyssey cluster supported by the FAS Division of Science, Research Computing Group at Harvard University. Post-processing analysis of simulation data was run on the Comet cluster hosted at SDSC, making use of the Extreme Science and Engineering Discovery Environment (XSEDE), which is supported by National Science Foundation grant number ACI-1053575.

%%%%%%%%%%%%%%%%%%%% REFERENCES %%%%%%%%%%%%%%%%%%

\bibliographystyle{apj}
\bibliography{imf}

%%%%%%%%%%%%%%%%% APPENDICES %%%%%%%%%%%%%%%%%%%%%

\appendix
%%\vspace{.14in}

\section{Input \& output IMF relations}
For each simulation studied, Table \ref{tab:relations} shows the input relations constructed based on the indicated star-formation quantity and the fit to the resulting log($\alpha_{\rm IMF}$)-log($\sigma_{*}$) output relation. The output relations listed under 0.5$R^{p}_{1/2}$ correspond to the relations shown in Figure \ref{fig:imf} and Figure \ref{fig:res_var}. 

\input{relations-table}

\label{lastpage}
\end{document}

%% file: illustris-table.tex
\begin{table*}

\centering
\tabletypesize{\scriptsize}
\caption{Simulation parameters}
\label{tab:illustris}
\tablewidth{0pt}

\begin{tabular}{lcccccccc}

\hline
Simulation  &  Volume &  $N_{\mathrm{snapshots}}$ & $N_{\mathrm{DM}}$ & $m_{\mathrm{DM}}$  & $\epsilon_{\mathrm{b}}$ & $\epsilon_{\mathrm{DM}}$ & $N_{\mathrm{galaxies}}$ & selected $N_{\mathrm{galaxies}}$ \\ 
& [(Mpc/$h$)$^{3}$] &  & & [M$_{\odot}$] & [kpc] & [ckpc] & [at $z = 0$] & [at $z = 0$]  \\ (1) & (2) & (3) & (4) & (5) & (6) & (7) & (8) & (9) \\

\hline
\hline

Illustris-1	&	75$^{3}$	&	134	&	1820$^{3}$	&	6.3 $\times$ 10$^{6}$ & 0.7 & 1.4 & 4366546 & 371   \\
Illustris-2	&	75$^{3}$	&	136	&	910$^{3}$	&	5.0  $\times$ 10$^{7}$ & 1.4 & 2.8 & 689785 & 229	\\
Illustris-3	&	75$^{3}$	&	136	&	455$^{3}$	&	4.0  $\times$ 10$^{8}$ & 2.8 & 5.7 & 121209 & 103	\\
No feedback	&	40$^{3}$	&	31	&	320$^{3}$	&	1.7  $\times$ 10$^{8}$ & 1.8 & 3.6 &  38363& 178$^{ \dagger}$	\\
Winds only	&	40$^{3}$	&	31	&	320$^{3}$	&	1.7  $\times$ 10$^{8}$ & 1.8 & 3.6 & 37121 &  41$^{ \dagger}$ \\	
IMF-Salpeter	&	40$^{3}$	&	31	&	320$^{3}$	&	1.7  $\times$ 10$^{8}$ & 1.8 & 3.6 &  38375& 23	\\
IMF-Spiniello	&	40$^{3}$	&	31	&	320$^{3}$	&	1.7  $\times$ 10$^{8}$ & 1.8 & 3.6 & 37638 &  31 \\

\hline

\multicolumn{9}{l}{(1) Simulation name; (2) Volume of box where the hubble constant is $h$ = 0.704 [100 km/s/Mpc];} \\

\multicolumn{9}{l}{(3) Number of snapshots produced; (4) Number of dark matter particles; (5) Mass of each dark matter particle; }  \\
\multicolumn{9}{l}{(6) Baryonic gravitational softening length at $z = 0$; (7) Dark matter gravitational softening length in comoving kpc; } \\
\multicolumn{9}{l}{(8) Number of galaxies identified at $z = 0$; (9) Number of massive ($M_{*}$~\textgreater ~10$^{10}$ M$_{\odot}$, $\sigma_{*}$ \textgreater ~150 km s$^{-1}$),} \\ 
\multicolumn{9}{l}{~quiescent (sSFR  \textless ~10$^{-11}$ yr$^{-1}$) galaxies at $z = 0$, ~$^{\dagger}$ no sSFR cut } \\

\end{tabular}

\bigskip

\end{table*}

%%% add details of other simulations used in study to this table

%% file: relations-table.tex
\begin{table*}[h]

\tabletypesize{\scriptsize}
\caption{IMF Relations}
\label{tab:relations}
\tablewidth{0pt}

\def\arraystretch{1.5}
\begin{tabular}{lclll}

\hline
Simulation & Quantity  &  Input relation &  \multicolumn{2}{l}{Output relation}  \\ 
\hline
&   &  & $R_{all}$ & 0.5$R^{p}_{1/2}$ \\ (1) & (2) & (3) & (4) & (5)  \\

\hline
\hline

Illustris-1 & $\sigma_{*}$ &	log($\alpha_{\rm IMF}$) = 1.05$\times$log($\sigma_{*}$) - 2.71 & log($\alpha_{\rm IMF}$) = 0.25$\times$log($\sigma_{*}$) - 0.76 & log($\alpha_{\rm IMF}$) = 0.42$\times$log($\sigma_{*}$) - 1.12 \\

&  & log($\alpha_{\rm IMF}$) = 3.5$\times$log($\sigma_{*}$) - 7.84 & log($\alpha_{\rm IMF}$) = 0.52$\times$log($\sigma_{*}$) - 1.31 & log($\alpha_{\rm IMF}$) = 0.96$\times$log($\sigma_{*}$) - 2.26 \\

& $\sigma_{\rm birth}$	 &	log($\alpha_{\rm IMF}$) = 1.05$\times$log($\sigma_{\rm birth}$) - 2.71	&	log($\alpha_{\rm IMF}$) = 0.21$\times$log($\sigma_{*}$) - 0.68	&	log($\alpha_{\rm IMF}$) = 0.38$\times$log($\sigma_{*}$) - 1.03 \\

 & 	&	log($\alpha_{\rm IMF}$) = 3.7$\times$log($\sigma_{\rm birth}$) - 8.99	&	log($\alpha_{\rm IMF}$) = 0.51$\times$log($\sigma_{*}$) - 1.30	&	log($\alpha_{\rm IMF}$) = 0.97$\times$log($\sigma_{*}$) - 2.29	\\

 & $\mathrm{[M/H}]$ 	&	log($\alpha_{\rm IMF}$) = 1$\times$[M/H] + 0.07	&	log($\alpha_{\rm IMF}$) = -0.12$\times$log($\sigma_{*}$) + 0.26	&	log($\alpha_{\rm IMF}$) = -0.12$\times$log($\sigma_{*}$) + 0.39	\\

 &  	&	log($\alpha_{\rm IMF}$) = 4$\times$[M/H] - 0.02 	&	log($\alpha_{\rm IMF}$) = -0.13$\times$log($\sigma_{*}$) + 0.32	&	log($\alpha_{\rm IMF}$) = -0.20$\times$log($\sigma_{*}$) + 0.65	\\

& $\sigma_{\rm gas}$	&	log($\alpha_{\rm IMF}$) =	1.05$\times$log($\sigma_{\rm gas}$) - 2.46 &	log($\alpha_{\rm IMF}$) =	0.24$\times$log($\sigma_{*}$) - 0.73&	log($\alpha_{\rm IMF}$) = 0.41$\times$log($\sigma_{*}$) - 1.10 \\	
 &  	&	log($\alpha_{\rm IMF}$) = 4.3$\times$log($\sigma_{\rm gas}$) - 9.38 	&	log($\alpha_{\rm IMF}$) =	0.51$\times$log($\sigma_{*}$) - 1.31 &	log($\alpha_{\rm IMF}$) = 0.96$\times$log($\sigma_{*}$) - 2.27	\\
& SFR	&	log($\alpha_{\rm IMF}$) = -0.15$\times$log(SFR) + 0.10	&	log($\alpha_{\rm IMF}$) = -0.12$\times$log($\sigma_{*}$) + 0.18	&	log($\alpha_{\rm IMF}$) =  -0.21$\times$log($\sigma_{*}$) + 0.34\\
 &  	&	log($\alpha_{\rm IMF}$) = -0.7$\times$log(SFR) + 1.32	&	log($\alpha_{\rm IMF}$) = -0.78$\times$log($\sigma_{*}$) + 2.01 	&	log($\alpha_{\rm IMF}$) = -1.03$\times$log($\sigma_{*}$) + 2.49	\\
\\

\hline

Illustris-2 & $\sigma_{\rm birth}$	 &	log($\alpha_{\rm IMF}$) = 1.05$\times$log($\sigma_{\rm birth}$) - 2.71 &	log($\alpha_{\rm IMF}$) = 0.22$\times$log($\sigma_{*}$) - 0.69	&	log($\alpha_{\rm IMF}$) = 0.42$\times$log($\sigma_{*}$) - 1.12  \\
 & 	&	log($\alpha_{\rm IMF}$) = 3.7$\times$log($\sigma_{\rm birth}$) - 8.99	&	log($\alpha_{\rm IMF}$) = 0.49$\times$log($\sigma_{*}$) - 1.25	&	log($\alpha_{\rm IMF}$) =	1.09$\times$log($\sigma_{*}$) - 2.54\\
\\

\hline

Illustris-3 & $\sigma_{\rm birth}$	 &	log($\alpha_{\rm IMF}$) = 1.05$\times$log($\sigma_{\rm birth}$) - 2.71 &	log($\alpha_{\rm IMF}$) = 0.21$\times$log($\sigma_{*}$) - 0.66	&	log($\alpha_{\rm IMF}$) = 0.38$\times$log($\sigma_{*}$) - 1.05   \\
 & 	&	log($\alpha_{\rm IMF}$) = 3.7$\times$log($\sigma_{\rm birth}$) - 8.99	& log($\alpha_{\rm IMF}$) = 0.47$\times$log($\sigma_{*}$) - 1.22	&	log($\alpha_{\rm IMF}$) =	1.05$\times$log($\sigma_{*}$) - 2.48\\
\\

\hline

No feedback & $\sigma_{\rm birth}$	 &	log($\alpha_{\rm IMF}$) = 1.05$\times$log($\sigma_{\rm birth}$) - 2.71 &	log($\alpha_{\rm IMF}$) = 0.20$\times$log($\sigma_{*}$) - 0.65	&	log($\alpha_{\rm IMF}$) = 0.37$\times$log($\sigma_{*}$) - 1.04  \\
 &  	&	log($\alpha_{\rm IMF}$) = 3.7$\times$log($\sigma_{\rm birth}$) - 8.99	&	log($\alpha_{\rm IMF}$) = 0.47$\times$log($\sigma_{*}$) - 1.25	&	log($\alpha_{\rm IMF}$) = 1.01$\times$log($\sigma_{*}$) - 2.44	\\
\\

\hline

Winds only & $\sigma_{\rm birth}$	 &	log($\alpha_{\rm IMF}$) = 1.05$\times$log($\sigma_{\rm birth}$) - 2.50 	&	log($\alpha_{\rm IMF}$) =	0.72$\times$log($\sigma_{*}$) - 1.71 &	log($\alpha_{\rm IMF}$) =  1.04$\times$log($\sigma_{*}$) - 2.42 \\
 &  	&	log($\alpha_{\rm IMF}$) = 3.7$\times$log($\sigma_{\rm birth}$) - 8.99	&	log($\alpha_{\rm IMF}$) = 1.17$\times$log($\sigma_{*}$) - 2.77	&	log($\alpha_{\rm IMF}$) =	2.19$\times$log($\sigma_{*}$) - 5.09 \\
\\

\hline

IMF-Salpeter & $\sigma_{\rm birth}$	 &	log($\alpha_{\rm IMF}$) = 1.05$\times$log($\sigma_{\rm birth}$) - 2.71 &	log($\alpha_{\rm IMF}$) = 0.13$\times$log($\sigma_{*}$) - 0.50	&	log($\alpha_{\rm IMF}$) = 0.20$\times$log($\sigma_{*}$) - 0.65  \\
 &  	&	log($\alpha_{\rm IMF}$) = 3.7$\times$log($\sigma_{\rm birth}$) - 8.99	&	log($\alpha_{\rm IMF}$) = 0.34$\times$log($\sigma_{*}$) - 0.95	&	log($\alpha_{\rm IMF}$) = 0.56$\times$log($\sigma_{*}$) - 1.44	\\
\\

\hline

IMF-Spiniello & $\sigma_{\rm birth}$	 &	log($\alpha_{\rm IMF}$) = 1.05$\times$log($\sigma_{\rm birth}$) - 2.71 	&	log($\alpha_{\rm IMF}$) =	0.14$\times$log($\sigma_{*}$) - 0.52 &	log($\alpha_{\rm IMF}$) =  0.26$\times$log($\sigma_{*}$) - 0.77 \\
 &  	&	log($\alpha_{\rm IMF}$) = 3.7$\times$log($\sigma_{\rm birth}$) - 8.99	&	log($\alpha_{\rm IMF}$) = 0.34$\times$log($\sigma_{*}$) - 0.91	&	log($\alpha_{\rm IMF}$) =	0.70$\times$log($\sigma_{*}$) - 1.69 \\

\hline
\hline

\multicolumn{5}{l}{(1) Simulation name; (2) Star-formation quantity; (3) Applied input relation;} \\
\multicolumn{5}{l}{(4) Fit to output relation with input relation applied to all star particles belonging to each galaxy;} \\
\multicolumn{5}{l}{(5) Fit to output relation with input relation applied to all star particles within 0.5$R^{p}_{1/2}$.}

\end{tabular}

\end{table*}